\documentclass[aps,twocolumn,prb,showpacs,nofootinbib]{revtex4-1}

\usepackage[english]{babel}
\usepackage{mathtools}
\usepackage{amsmath}
\usepackage{amsfonts}
\usepackage{amssymb}
\usepackage{graphicx,grffile}
\usepackage{bm}
\usepackage[usenames,dvipsnames]{xcolor}

\usepackage[draft,markup=default,authormarkuptext=name,
 commentmarkup=uwave]{changes}
\definechangesauthor[name={Win},color={Green}]{w}
\definechangesauthor[name={Andreas},color={orange}]{ak}
\definechangesauthor[name={Ara},color={blue}]{as}
\definechangesauthor[name={Ilya},color={magenta}]{i}
\usepackage{fixmetodonotes}

\usepackage{tikz,pgfplots}
\usetikzlibrary{matrix,arrows,decorations.markings}
\tikzset{->-/.style={decoration={markings,
  mark=at position .5 with {\arrow{>}}},postaction={decorate}},
  -<-/.style={decoration={markings,
    mark=at position .5 with {\arrow{<}}},postaction={decorate}}}

\usepackage{hyperref}

\newcommand{\be}{\begin{equation}}
\newcommand{\ee}{\end{equation}}
\newcommand{\bes}{\begin{equation}\begin{split}}
\newcommand{\ees}{\end{split}\end{equation}}
\newcommand{\bea}{\begin{eqnarray}}
\newcommand{\eea}{\end{eqnarray}}

\newcommand{\rez}[1]{\frac{1}{#1}}

\def\beq{\begin{equation}}
\def\eeq{\end{equation}}
\def\bea{\begin{eqnarray}}
\def\eea{\end{eqnarray}}

\begin{document}

\title{Random network models with variable disorder of geometry}

\author{A. Kl\"umper}
\affiliation{Bergische Universit\"at Wuppertal, Gau\ss stra\ss e 20, 42119
Wuppertal, Germany}

\author{W. Nuding}
\affiliation{Bergische Universit\"at Wuppertal, Gau\ss stra\ss e 20, 42119
Wuppertal, Germany}

\author{A. Sedrakyan}
\affiliation{Yerevan Physics Institute, Br. Alikhanian 2, Yerevan 36,
Armenia}
\affiliation{International Institute of Physics, Natal, Brazil}

\begin{abstract}		
Recently it was shown (I.~A.~Gruzberg, A. Kl\"umper, W. Nuding and A. Sedrakyan,
Phys.~Rev.~B 95, 125414 (2017)) that taking into account random positions of
scattering nodes in the network model with $U(1)$ phase disorder yields a
localization length exponent $2.37 \pm 0.011$ for plateau transitions in the
integer quantum Hall effect. This is in striking agreement with the experimental
value of $2.38 \pm 0.06$. Randomness of the network was modeled by replacing
standard scattering nodes of a regular network by pure tunneling
resp.~reflection with probability $p$ where the particular value $p=1/3$ was
chosen.  Here we investigate the role played by the strength of the geometric
disorder, i.e. the value of $p$.  We consider random networks with arbitrary
probability $0 <p<1/2$ for extreme cases and show the presence of a line of
critical points with varying localization length indices having a minimum
located at $p=1/3$.
\end{abstract}

\date{June 27, 2019}


\maketitle

{\large {\it Introduction}} The physics of plateau transitions in the Integer
Quantum Hall Effect (IQHE) poses a crucial condensed matter problem
potentially necessitating a new understanding of quantum criticality.  It
relates not only to chiral systems where time reversal symmetry (TRS) is
broken, but also to topological insulators (TI) with TRS. This transition is
an example of a metal-insulator transition in two dimension where TRS is
broken due to the presence of a magnetic field.  In their seminal paper
\cite{Chalker-Percolation-1988} Chalker and Coddington suggested a
phenomenological model (CC model) for edge excitations in magnetic fields,
where the disorder potential creates a scattering network based on quantum
tunneling between Fermi levels of neighbor Fermi ``puddles'' in the ground
state. For simplicity, the authors suggested that the scattering nodes in the
landscape of the random potential are disposed regularly, while the
information about randomness is coded in random $U(1)$ phases associated with
the links of the network. During the last 30 years there were huge activities
\cite{Huckestein-Scaling-1995, Huckestein-1992, Huckestein-1994,
  Lee-Quantum-1993,Ludwig-Integer-1994, Bhatt-1996, Chamon-Instability-1996,
  Lee-Effects-1996, Klesse-Universal-1995, Li-Scaling-2005,
  Zirnbauer-Conformal-1999, Zirnbauer-Riemannian-1996, Zirnbauer-Towards-1994,
  Galstyan-Raikh-1997, Cain-Romer-Raikh-2003, Mkhitaryan-Raikh-2009,
  Mkhitaryan-Kagalovsky-Raikh-2009, Song-Prodan-2013} in understanding the CC
model, its continuum limit and links to conformal field theories
\cite{Bhaseen-Towards-2000,Tsvelik-Wave-2001,Leclair2001}. A Spin Hall analog
of the CC model was formulated \cite{Kagalovsky-Quantum-1999} and investigated
in \cite{Gruzberg-Exact-1999}. It appeared, however, that the position of
scattering nodes on a regular lattice looses an essential part of the
randomness of the potential. Numerical calculations of the Lyapunov exponent
in the CC model give a localization length index $\nu= 2.56 \pm 0.011$
\cite{Slevin-Critical-2009,Amado-Numerical-2011,Slevin-Finite-2012,
  Obuse-Finite-2012, Dahlhaus-Quantum-2011, Nuding-Localization-2015}, which
is well separated from the experimental value $\nu =2.38 \pm 0.06$
\cite{Wei-Current-1994,Li-Scaling-2005,Li-Scaling-2009}. {Recently,
  alternatives to the CC model approach give values $\nu \simeq 2.58$ in
  Ref.~[\onlinecite{Vojta-2019}] and $\nu = 2.48 \pm 0.02$ in
  [\onlinecite{Bhatt-2019}] with only the latter being just compatible with
  the experimental result.}

\begin{figure}[h]
	\centering
	\includegraphics[width=0.7\columnwidth, angle=0]{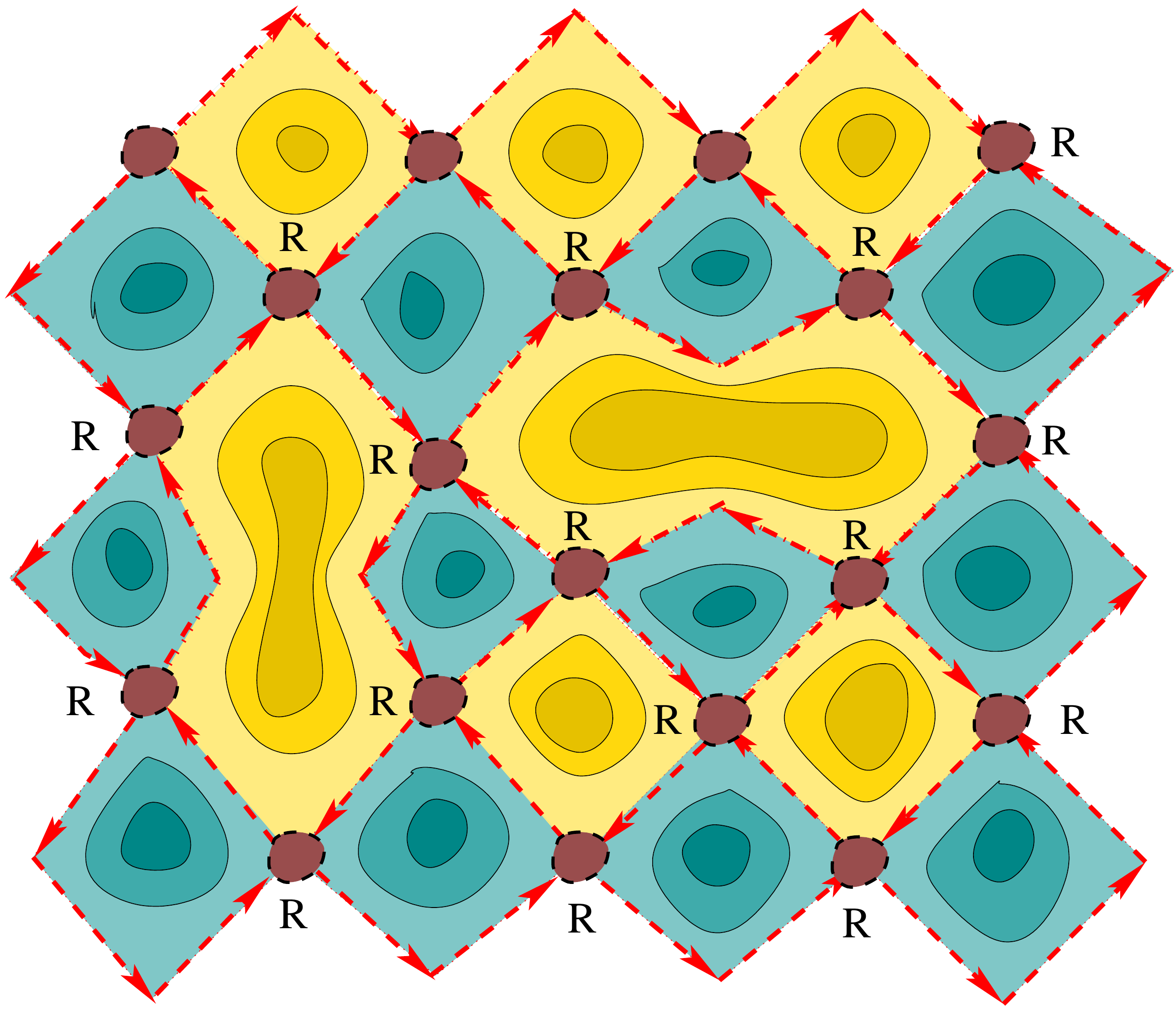}
	\caption{Modified CC network with two ``open'' nodes, one in the vertical
		and one in the horizontal direction.}
	\label{fig:random-pot-1}
\end{figure}

\begin{figure}[t]
	\centering
	\includegraphics[
	width=0.75\columnwidth, angle=0]{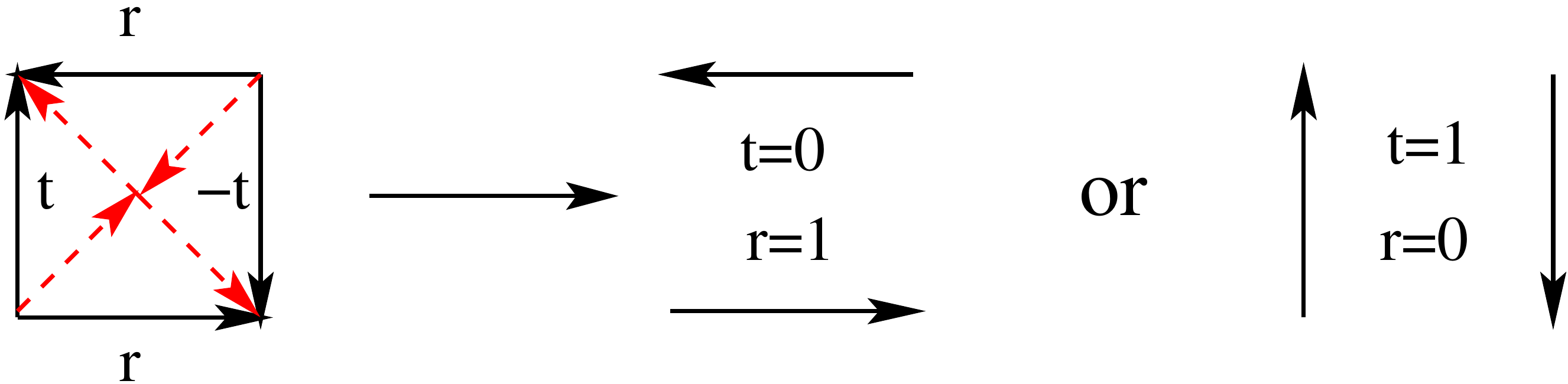}
	\vskip 2mm
	\includegraphics[
	width=0.9\columnwidth,
%        angle=0]{/home/kluemper/Dropbox/modified CC-lattice/1_3-1_3-1_3 model/figs/CC-modification.pdf}
        angle=0]{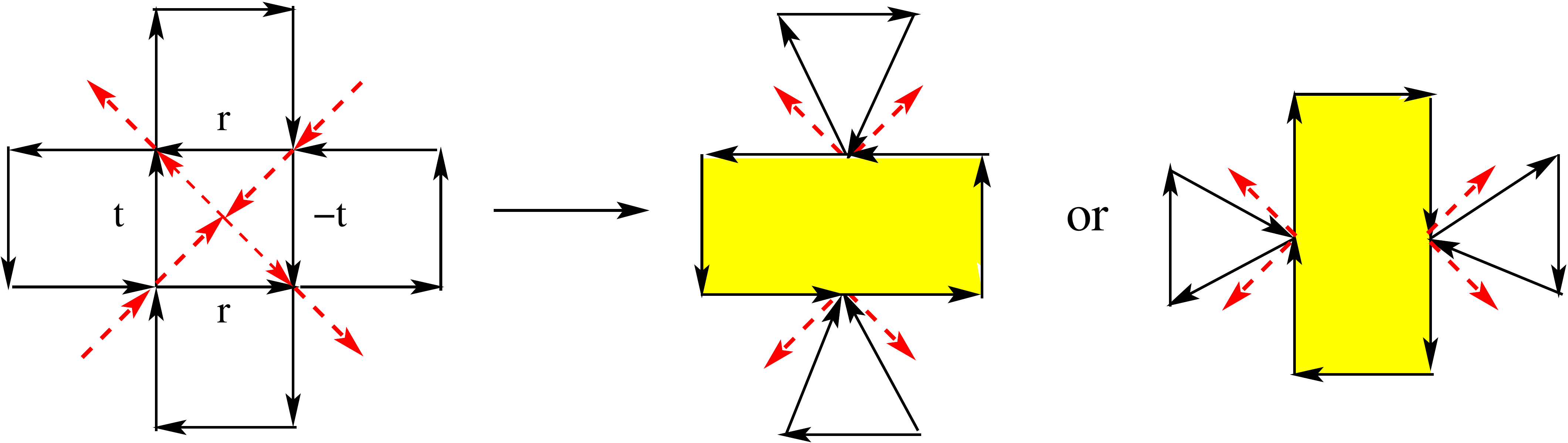}
	\caption{Top: Graphical illustration of the opening of a node. Bottom:
          The resulting modifications of the medial lattice.}
	\label{fig:open-nodes}
	%  \vskip -4mm
\end{figure}

\begin{figure}[h]
	\centering
	\includegraphics[width=0.7\columnwidth,angle=0]{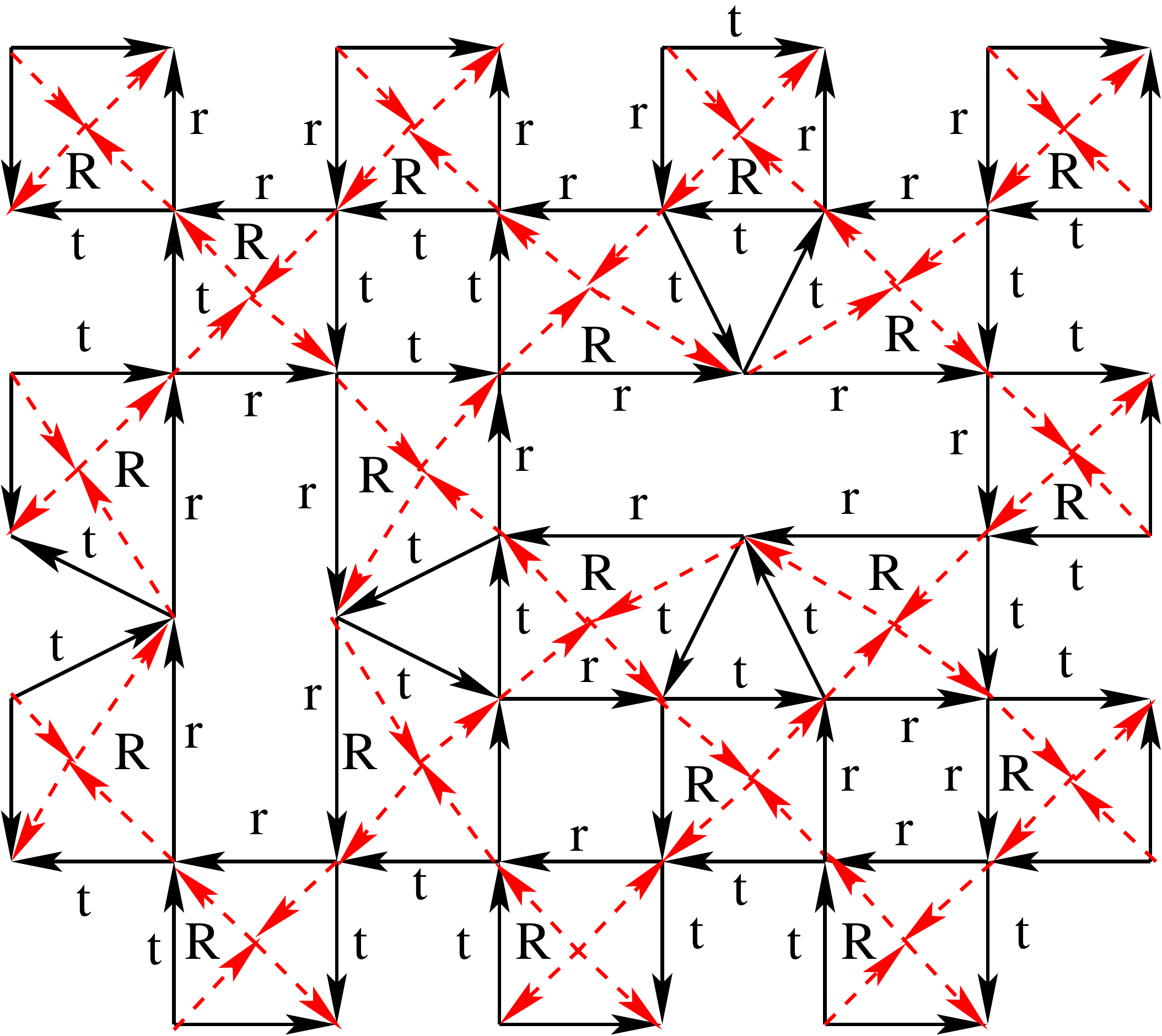}
	\caption{The modified medial lattice corresponding to the network
          shown in Fig.~\ref{fig:random-pot-1}. We see two hexagons
          (vertically and horizontally oriented) {appearing in
            neighborhood with two triangles each}. Red dotted lines here
          correspond to red dotted lines in Fig.1.
        }
	\label{fig:random-lattice-1}
\end{figure}

The discrepancy between the experimental value of $\nu$ and the
  CC model prediction may be due to the importance of electron-electron
  interactions studied in papers
  ~[\onlinecite{Polyakov-Shklovskii-1993},\onlinecite{Pruisken-Baranov-1995},\onlinecite{Pruisken-Burmistrov-1995},\onlinecite{Girvin-Chalker-2000},
    \onlinecite{Burmistrov-2011}].  However another solution of this problem
was proposed in the paper \cite{GruzbergNudingKluemperSedrakyan2017}, based on
the observation that randomness of the relative positions of nearest neighbor
scattering nodes has to be taken into account. For a depiction of a disorder
potential with non-regular positioned saddle points see
Fig.~\ref{fig:random-pot-1}. This randomness of the network leads to the
appearance of curvature in 2d space and may be regarded as the induction of
quenched 2d gravity, which changes the universality class of the problem. In
order to generate disordered networks in the transfer matrix formalism a
new model was formulated, where the regular scattering 
with $S$-matrix 
$S= \left( \begin{array} {cc} r & t\\ -t & r \\ \end{array} \right)$
at the saddle points is randomly replaced by two other extreme events. Here,
the $S$-matrix takes the form of complete reflection, $(t,r)=(0,1)$, with
probability $p_1$ or the form of complete tunneling, $(t,r)=(1,0)$, with
probability $p_2$ as presented in Fig.~\ref{fig:open-nodes}. The probability of
regular scattering events is $p_3=1-p_1-p_2$. The two extreme scattering
events eliminate links in the scattering network. They perform a kind of
``surgery'' to a flat network where $n$-faces with $n=3, 5, 6, \cdots$ appear
in the lattice. Examples of such ``surgery'' are presented in
Fig.~\ref{fig:open-nodes}.
Following this procedure we can formulate a hopping model of fermions on a
random Manhattan lattice (ML), as is presented in
Fig.~\ref{fig:random-lattice-1}, which corresponds to the landscape of the
potential presented in Fig.~\ref{fig:random-pot-1}.

The appearance of $n$-faces in the ML means, that our 2d geometry is not flat
anymore and contains local Gaussian curvature $R_n= \frac{\pi}{2}(4-n)$ for
each $n$-face with $n\not=4$. {This is the discrete analog of the Gauss
  curvature integrated over a face $\int_{face} R \sqrt{g}\, d^2 \xi$.} Hence,
the average over randomness of the saddle points leads to the average over all
configurations of the curved space~[\onlinecite{Ambjorn-Sedrakyan-2013},
  \onlinecite{Ambjorn-Sedrakyan-2015}], with yet to be determined functional
measure.  The field, which is characterizing different surfaces and by use of
which one can ensure reparametrization invariance of the model is the metric,
while {the} corresponding theory is 2d gravity. This indicates that we have a
non-critical string model, where all physical variables should be invariant
under arbitrary coordinate transformations. The appearance of this new field
and symmetry is the {reason for the changes of the critical indices of the
  flat problem. And, as it appeared
  [\onlinecite{GruzbergNudingKluemperSedrakyan2017}], by taking equal
  probability 1/3 for each of the three nodes (complete reflection, complete
  transmission, regular scattering), the localization length index becomes
  $\nu=2.37 \pm 0.011$, very close to the experimental value.  In passing, we
  like to remark that recently the problem of the fractional quantum Hall
  effect on arbitrary gravitational background has attracted considerable
  interest
  \cite{Can-Laskin-Wiegmann-2014,Can-Laskin-Wiegmann-2015,Laskin-Can-Wiegmann-2015}.

In papers Ref.~[\onlinecite{Vojta-2019}, \onlinecite{Bhatt-2019}] the regular
tight-binding lattice model in a magnetic field with random site energies was
numerically analyzed. In [\onlinecite{Vojta-2019}] the authors considered the
one particle Green's function and got $\nu=2.58(4)$ for the correlation length
index, while in [\onlinecite{Bhatt-2019}] the density of states around zero
energy was analyzed yielding $\nu=2.48\pm 0.02$.  The first paper confirms
the result for the standard CC model.
 Our modified CC model differs essentially from this model 
   because it contains information about the geometry of filled Landau levels,
   which form ``lakes'' in a random potential background.  Assuming that the
 numerical analysis of both models was done on sufficiently large lattices we
 have to conclude that the models with different critical
   indices belong to different universality classes.
  
Another important question appearing here is the validity of the Harris
criterion \cite{Harris-1974, Chayes-Chayes-1986}. According to it, for $d
\nu>2$
with $d$ being the spatial dimension, any new disorder cannot change the
critical index $\nu$ of the system. For the CC model we find $\nu \sim 2.56$
\cite{Slevin-Critical-2009,Amado-Numerical-2011,Slevin-Finite-2012,
  Obuse-Finite-2012, Dahlhaus-Quantum-2011, Nuding-Localization-2015} so the
above condition is fulfilled. Therefore one naively may expect, that disorder
connected with randomness of the network cannot change the CC model
localization length index.
  
The fundamental arguments leading to Harris criterion are based on the
following observations, see for instance \cite{Vojta-2013}: We consider a
system at some temperature $T$ close to the critical temperature $T_c$ of the
ordered bulk. We then divide the space into correlated blocks of the size of
the correlation length $\xi(T)$. Each block $i$ has its own realization of
disorder and has a corresponding transition temperature $T_{c,i}$.  If the
deviation of the critical temperatures, thanks to the central limit theorem of
the order $\Delta{T}_i\sim \xi^{-d/2}$, is smaller than the distance of the
actual temperature $T$ from the critical point $T-T_c \sim \xi^{-1/\nu}$, then
a uniform phase transition happens and the disorder is irrelevant. In the
other case, different blocks may stay on different sides of the critical point
$T_c$ and far from it which will change the critical behavior. This is the
case of geometric disorder involving a finite fraction of extreme nodes with
$(t,r)=(0,1)$ or $(t,r)=(1,0)$ deviating considerably from the CC critical
point $r_c=t_c=1/\sqrt{2}$. In general, it appears questionable
  if the RG perturbative reasoning applies to strong disorder. Investigations
  concerning this issue are available \cite{Chayes-Chayes-1986}. In summary,
  the presented arguments can not be considered as proof and the influence of
  geometric disorder on the applicability of Harris' criterion needs further
  investigation.
 
A natural question that appears is, what is the meaning of probabilities
$p_i,\; i=1, 2, 3$ and do the critical indices of the model depend on them?
In this paper we consider the model with singular blocks (see
Fig.~\ref{fig:open-nodes}b and Fig.~\ref{fig:open-nodes}c) appearing in the
network with equal probabilities $p_1=p_2=p$, while the regular scattering has
the probability $1-2 p$. It is clear, that $p\leq 1/2$.

{{\it Construction and simulation of random networks.}}\\ For the calculation
of the correlation length index of our model we used a variant of the
transfer-matrix method formulated in Refs.
[\onlinecite{MacKinnon-One-Parameter-1981},
  \onlinecite{MacKinnon-The-scaling-1983}] and further developed in Ref.
[\onlinecite{Chalker-Percolation-1988, Amado-Numerical-2011}].  We calculate
the product
\begin{align}
{\cal T}_L=\prod_{j=1}^L T_{1j} U_{1j}T_{2j} U_{2j}
\end{align}
of layers of transfer matrices $M_1 U_{1j}M_2 U_{2j}$ corresponding to two
columns $T_{1j}$ and $T_{2j}$ of vertical sequences of $2\times2$ scattering nodes,
\begin{equation} \label{M1}
T_{1j}= \begin{tikzpicture}[baseline=(current bounding box.center), ultra thick, loosely dotted]
\matrix(T1)[matrix of math nodes,
nodes in empty cells,
right delimiter={)},
left delimiter={(}]
{
	T^{1}_{\alpha_1} & 0   & 	& 0  	\\
	0   &	T^{1}_{\alpha_2}	&  	& 		\\
	& 		& 	& 0 	\\
	0 	& 		& 0	& T^{1}_{\alpha_M}	\\
};
\draw (T1-2-2)--(T1-4-4);
\draw (T1-1-2)--(T1-1-4);
\draw (T1-4-1)--(T1-4-3);
\draw (T1-2-1)--(T1-4-1);
\draw (T1-1-4)--(T1-3-4);
\draw (T1-1-2)--(T1-3-4);
\draw (T1-2-1)--(T1-4-3);
\end{tikzpicture}
\end{equation}
and
\begin{equation}\label{M2}
T_{2j}=\begin{tikzpicture}[baseline=(current bounding box.center),
ultra thick, loosely dotted]
\matrix(T2)[matrix of math nodes,nodes in empty cells,
right delimiter={)},left delimiter={(}] {
	\big[T^2_{\alpha_1}\big]_{22}	&	0  	&		&	0	&	\big[T^2_{\alpha_1}\big]_{21}		\\
	0					&	T^2_{\alpha_2}	&		&			&	0					\\
	& 		&		&			&						\\
	0      		&			&		&	T^2_{\alpha_{M-1}}	&	0		\\
	\big[T^2_{\alpha_1}\big]_{12}	&	0		&		& 0		&	\big[T^2_{\alpha_1}\big]_{11}	\\
};
\draw (T2-1-2)--(T2-1-4);
\draw (T2-1-2)--(T2-4-5);
\draw (T2-2-1)--(T2-4-1);
\draw (T2-2-1)--(T2-5-4);
\draw (T2-2-2)--(T2-4-4);
\draw (T2-2-5)--(T2-4-5);
\draw (T2-5-2)--(T2-5-4);
\end{tikzpicture}.
\end{equation}
Here the index $\alpha_i=1, 2, 3$ should be randomly fixed; $\alpha=1$ with
probability $1-2 p$ for regular scatterings
\begin{equation}
T^1_1=\begin{pmatrix}
1/t & r/t \\
r/t & 1/t
\end{pmatrix},\;\;
%\qquad \text{and} \qquad
T^2_1=\begin{pmatrix}
1/r & t/r \\
t/r & 1/r
\end{pmatrix},
\end{equation}
or $\alpha=2,3$ with probability $p$, for ``surgery'' operations,
i.e.~``extremal scatterings''
\begin{equation}
%\end{equation}
T^{1/2}_2=\begin{pmatrix}
1 & 0 \\
0 & 1
\end{pmatrix},\;
%\qquad \text{and} \qquad
T^{1/2}_3=\begin{pmatrix}
1/\epsilon & \sqrt{1-\epsilon^2}/\epsilon \\
\sqrt{1-\epsilon^2}/\epsilon & 1/\epsilon
\end{pmatrix}.
\end{equation}
The parameter $\epsilon$ here is a regularization parameter, which ideally
should be set to zero after the calculation of the Lyapunov exponent.

This choice of the transfer matrices corresponds to periodic boundary
conditions in the transverse direction. In other words, these transfer
matrices describe the random network model on a cylinder.

The $U$-matrices have a simple diagonal form with independent phase factors
$U_{nm}=\exp{(i\phi_n)}\,\delta_{nm}$ for $U=U_{1j}$ and
$U_{2j}$. The parameters $t$ and $r$ of the regular scattering are
the transmission and reflection amplitudes at each node and we parameterize
them as in the previous paper
[\onlinecite{GruzbergNudingKluemperSedrakyan2017}]
\begin{equation} \label{rt}
t=\rez{\sqrt{1+e^{2x}}} \qquad \text{and} \qquad r=\rez{\sqrt{1+e^{-2x}}}.
\end{equation}
The parameter $x$ corresponds to the Fermi energy measured from the Landau
band center scaled by the Landau band width. Following paper
[\onlinecite{GruzbergNudingKluemperSedrakyan2017}] we expect that the critical
point of the model of arbitrary $p$ is still given by the value $t_c^2=1/2$
{as for the regular nodes} corresponding to $x=0$. The phases $\phi_{n}$ are
random variables uniformly distributed in the range $[0,2\pi)$, reflecting
  that the phase of an electron approaching a saddle point of the random
  potential is arbitrary.

\begin{figure}[h]
	\centering
	\includegraphics{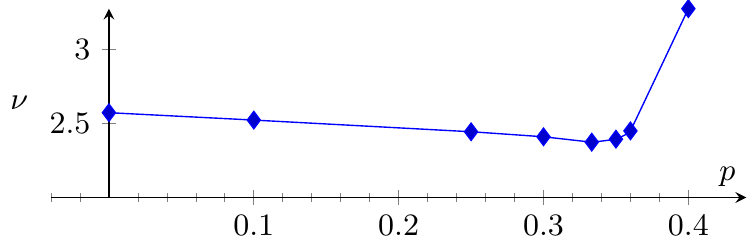}
	\caption{Localization length index $\nu_p$ versus probability $p$ of singular blocks}
	\label{fig:random-pot-2}
\end{figure}

\begin{figure}[h]
	\centering
	\includegraphics{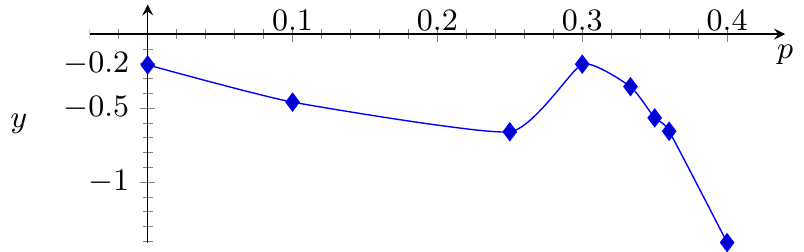}
	\caption{Subleading index versus probability $p$ of singular blocks}
	\label{fig:random-pot-3}
\end{figure}

\begin{figure}[h]
	\centering
	\includegraphics{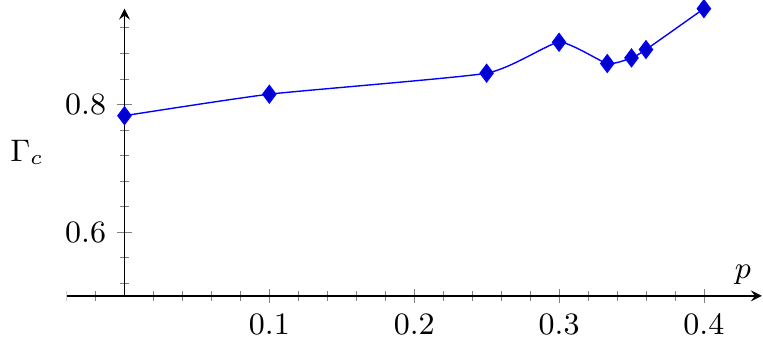}
	\caption{Coefficient $\Gamma_c=\pi (\alpha_0 - 2)$ related to the multifractal exponent $\alpha_0$
          versus probability p of singular blocks}
	\label{fig:random-pot-4}
\end{figure}

To extract the exponent $\nu$ for random networks, we numerically estimate
the Lyapunov exponent $\gamma$ defined as the smallest positive eigenvalue of
\bea 
	\label{LE} 
	\frac{1}{2L} \log[T_L T_L^\dagger]. 
\eea 
in the limit $L\to \infty$. In the standard transfer matrix method
one multiplies many transfer matrices for a single realization of disorder and
relies on the self-averaging property of Lyapunov exponents. This property in
the limit of infinite length of the sample is the subject of the
central-limit-type theorem for products of random matrices due to Oseledec.
\cite{Oseledec-A-Multiplicative-1968} The modification of 
Ref.~[\onlinecite{Amado-Numerical-2011}] that we use here, however, is
based on another central-limit-type theorem for products of random matrices
due to Tutubalin. \cite{Tutubalin-On-Limit-1965} This theorem states that the
Lyapunov exponents of products of a finite number of random matrices are
random numbers whose distribution approaches Gaussian for large sample
lengths.

These theorems allow us to simulate ensembles of {$N_r=624$} strips of height
$M$ (the number of nodes per column, varying from $20$ to $200$) {in the case
  of $q=1/3$} and length {$L=5\cdot10^6$.} {This is} equivalent
\cite{Amado-Numerical-2011} to the standard transfer matrix simulation of a
single sample of effective length $L_\text{eff} = N_r \times L >
{3\cdot10^9}$, exceeding the longest previously reported sample lengths.
{Moreover this method allows for an estimate of the precision of the
  calculated Lyapunov exponents by means of the standard deviation of those
  ensembles.}  The range of the parameter $x$ we have considered is $x\in
{[0,0.08]}$ which encodes deviations of $t$ from $t_c$. Then we fit all data
of the Lyapunov exponent for pairs of the parameters $(M,x)$ extracting the
localization index $\nu$. For each ensemble of the random network we check that
the histogram of the Lyapunov exponents is close to a Gaussian.

We use the so-called LU decomposition of transfer matrices
\cite{Nuding-Localization-2015}, because it {is} faster than the standard QR
decomposition approach.  Since $t$ and $r$ appear in the denominators of the
matrix elements of transfer matrices, making them zero is a singular
procedure, related to the disappearance of two horizontal channels upon
opening a node in the vertical direction (see Fig.~\ref{fig:open-nodes}).  To
overcome this difficulty, following
[\onlinecite{GruzbergNudingKluemperSedrakyan2017}] we take for every open
node either $t$ or $r$ to be equal to $\varepsilon \ll 1$.  It appears, that
the result for the Lyapunov exponent is unchanged within our error $10^{-3}$
in a range from $\varepsilon = 10^{-5}$ up to $\varepsilon$ to $10^{-7}$. For
even smaller $\varepsilon$ the results start changing again. This is to be
expected because the large differences of values in the entries of transfer
matrices cause numerical instabilities for the LU
decomposition. Interestingly, we found that the results for the Lyapunov
exponents for longer chains depend less on the value of $\epsilon$ than for
shorter chains. We have chosen $\varepsilon=10^{-6}$ for our calculations.

As usual, the Lyapunov exponent $\gamma$ is expected to have the
following finite-size scaling behavior:
\begin{align}
	\gamma M =\Gamma[M^{1/\nu}u_0(x), M^y u_1(x)]. 
\end{align}
Here $u_0(x)$ is the relevant field and $u_1(x)$ is the leading irrelevant
field. The relevant field vanishes at the critical point, and $y<0$.  The
fitting and the error analysis of our numerical data are described in the
appendices. The results of the analysis as functions of the disorder parameter
$p$ are presented in Fig.~\ref{fig:random-pot-2} for the localization length
exponent $\nu$, in Fig.~\ref{fig:random-pot-3} for the exponent $y$ of the
irrelevant field and in Fig.~\ref{fig:random-pot-4} for the parameter $\Gamma_c =
\pi (\alpha_0 - 2)$ related to the multifractal exponent $\alpha_0$. In table
\ref{table1} we present these results as numbers.

\begin{table}
	\begin{tabular}{|c||c|c|c|c|c|c|}
		\hline
%		\textbf{q1} & \textbf{q2} & \textbf{Γ0} & \textbf{ΔΓ0} & \textbf{ν} & \textbf{Δν} & \textbf{y} & \textbf{Δy} \\ \hline
		\textbf{$p$} & \textbf{$\Gamma_c$} &
                \textbf{$\Delta\Gamma_c$} & \textbf{$\nu$} &
                \textbf{$\Delta\nu$} & \textbf{$y$} & \textbf{$\Delta y$} \\ \hline
0           & 0.7823      & 0.05695      & 2.573      & 0.0145      & -0.2078    & 0.3744      \\
0.1         & 0.816       & 0.00595      & 2.523      & 0.0213      & -0.4592    & 0.1089      \\
0.25        & 0.8489      & 0.00295      & 2.444      & 0.017       & -0.6598    & 0.0527      \\
0.3         & 0.8974      & 0.07275      & 2.41       & 0.027       & -0.2028    & 0.0588      \\
1/3      & 0.864       & 0.864        & 2.374      & 0.0175      & -0.355     & 0.05        \\
0.35        & 0.8728      & 0.04895      & 2.394      & 0.015       & -0.5661    & 1.7         \\
0.36        & 0.8859      & 0.04395      & 2.45       & 0.0395      & -0.6562    & 1.9235      \\
0.4         & 0.95        & 0.00465      & 3.276      & 0.082       & -1.408  & 0.6487      \\ 
\hline
	\end{tabular}
\caption{Numerical values for the exponents $\nu$, $y$ and the multifractal
  parameter $\Gamma_c$ and their uncertainties. Different cases of the
  disorder parameter $p\in[0,1[$ are considered.}
\label{table1}
\end{table}

Fig.~\ref*{fig:random-pot-2} shows {an} interesting behavior of $\nu$ versus the
probability $p$.  We see that a minimum is achieved precisely at $p=1/3$ which
may very likely correspond to the plateau transitions in IQHE. The value $p=0$
gives $\nu=2.56$ for the Chalker-Coddington model, just as expected.  At
$p=1/2$, where we do not have regular scattering nodes at all the $x$ dependence
of $\gamma$ should disappear. Therefore, one can expect $\nu =\infty$, because
precisely in this situation the critical behavior of the Lyapunov exponent of
the form $x^\nu$ will produce zero.
As we see from Fig.~\ref*{fig:random-pot-2}, the index sharply increases close to
$p=1/2$.

{{\it Results and summary.}}\\ In summary, we have considered the possibility
that a certain type of geometric disorder, previously missing in the study of
the integer QH transition, changes its universality class. Our numerical
simulations support this idea.  We see that the random occurrence of singular
blocks in the network with some probability $p$ leads to a geometry with
curvature. The network model has a critical index $\nu_p$ that apparently
changes continuously with $p$, i.e.~it realizes a line of critical points with
different universality classes at different points. {The
  $p_1-p_2$ phase diagram of the model is presented in Fig.~\ref{phase-space},
  where the diagonal line from zero to $A$ is a line of critical points.}
\begin{figure}[t]
	\centering
	\includegraphics[width=0.5\columnwidth,angle=0]{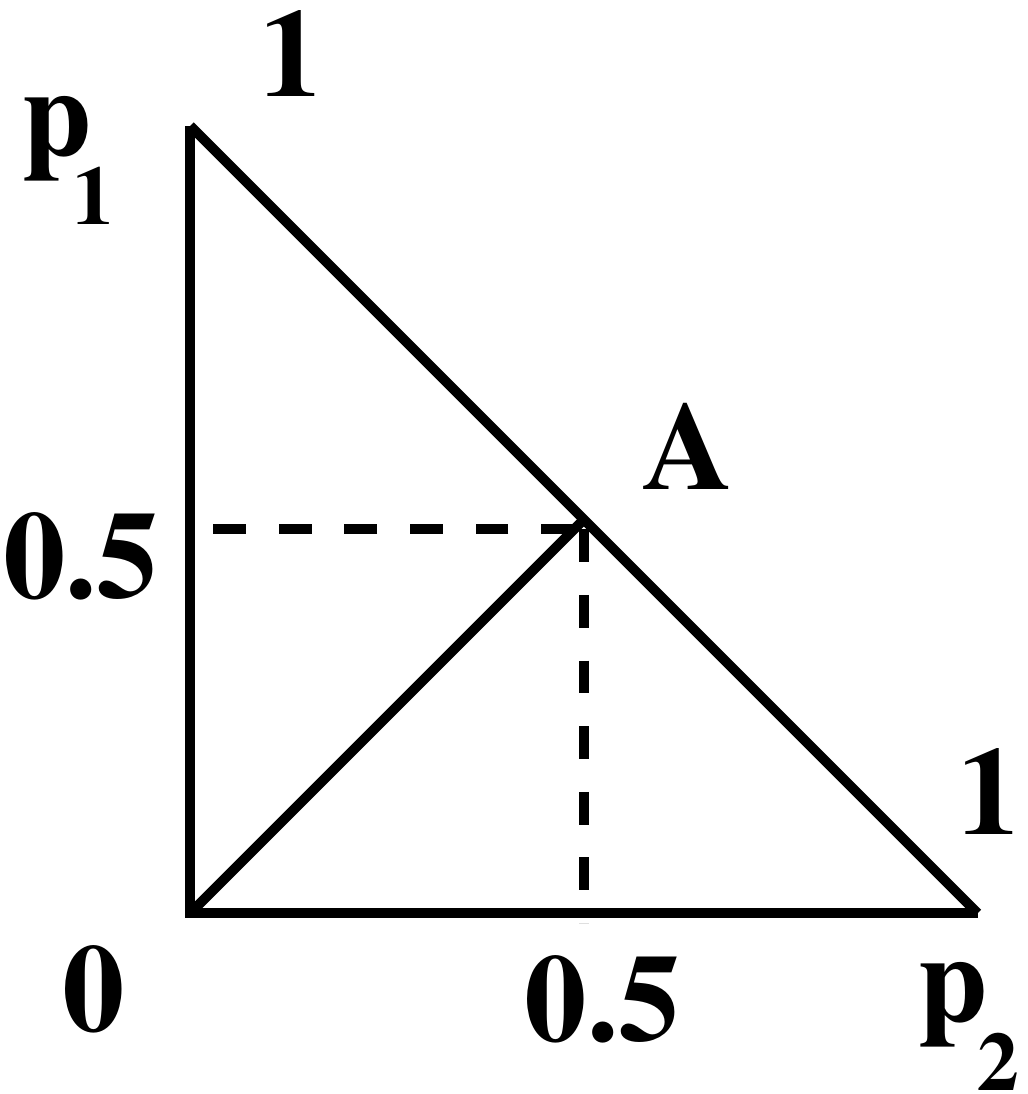}
	\caption{The $(p_1,p_2)$ phase diagram of the model
            ($p_1+p_2 \leq 1$). The segment $[0,A] $ is a line of critical
            points.}
	\label{phase-space}
\end{figure}

{The minimal value of $\nu$ at $p=1/3$ corresponds to the value expected for
  the exponent of the IQH transitions.  {The meaning of the other models as
    well as the meaning of the parameter $p$ remains an open question at the
    moment.}} {It would be not surprising, if the approaches
  presented in papers [\onlinecite{Vojta-2019}] and
  [\onlinecite{Bhatt-2019}] were related to different pameters $p$ in our
  model. }

\begin{acknowledgments}
A.~S. was supported by ARC grants 18T-1C153 and 18RF-039. A.~K. acknowledges
financial support by DFG. The authors are grateful to
  I.~A.~Gruzberg for many stimulating discussions and valuable comments. 
\end{acknowledgments}

\bibliography{RN2-bibliography}

\begin{thebibliography}{54}%
\makeatletter
\providecommand \@ifxundefined [1]{%
 \@ifx{#1\undefined}
}%
\providecommand \@ifnum [1]{%
 \ifnum #1\expandafter \@firstoftwo
 \else \expandafter \@secondoftwo
 \fi
}%
\providecommand \@ifx [1]{%
 \ifx #1\expandafter \@firstoftwo
 \else \expandafter \@secondoftwo
 \fi
}%
\providecommand \natexlab [1]{#1}%
\providecommand \enquote  [1]{``#1''}%
\providecommand \bibnamefont  [1]{#1}%
\providecommand \bibfnamefont [1]{#1}%
\providecommand \citenamefont [1]{#1}%
\providecommand \href@noop [0]{\@secondoftwo}%
\providecommand \href [0]{\begingroup \@sanitize@url \@href}%
\providecommand \@href[1]{\@@startlink{#1}\@@href}%
\providecommand \@@href[1]{\endgroup#1\@@endlink}%
\providecommand \@sanitize@url [0]{\catcode `\\12\catcode `\$12\catcode
  `\&12\catcode `\#12\catcode `\^12\catcode `\_12\catcode `\%12\relax}%
\providecommand \@@startlink[1]{}%
\providecommand \@@endlink[0]{}%
\providecommand \url  [0]{\begingroup\@sanitize@url \@url }%
\providecommand \@url [1]{\endgroup\@href {#1}{\urlprefix }}%
\providecommand \urlprefix  [0]{URL }%
\providecommand \Eprint [0]{\href }%
\providecommand \doibase [0]{http://dx.doi.org/}%
\providecommand \selectlanguage [0]{\@gobble}%
\providecommand \bibinfo  [0]{\@secondoftwo}%
\providecommand \bibfield  [0]{\@secondoftwo}%
\providecommand \translation [1]{[#1]}%
\providecommand \BibitemOpen [0]{}%
\providecommand \bibitemStop [0]{}%
\providecommand \bibitemNoStop [0]{.\EOS\space}%
\providecommand \EOS [0]{\spacefactor3000\relax}%
\providecommand \BibitemShut  [1]{\csname bibitem#1\endcsname}%
\let\auto@bib@innerbib\@empty
%</preamble>
\bibitem [{\citenamefont {{Chalker}}\ and\ \citenamefont
  {{Coddington}}(1988)}]{Chalker-Percolation-1988}%
  \BibitemOpen
  \bibfield  {author} {\bibinfo {author} {\bibfnamefont {J.~T.}\ \bibnamefont
  {{Chalker}}}\ and\ \bibinfo {author} {\bibfnamefont {P.~D.}\ \bibnamefont
  {{Coddington}}},\ }\href {\doibase 10.1088/0022-3719/21/14/008} {\bibfield
  {journal} {\bibinfo  {journal} {J. Phys. C}\ }\textbf {\bibinfo {volume}
  {21}},\ \bibinfo {pages} {2665} (\bibinfo {year} {1988})}\BibitemShut
  {NoStop}%
\bibitem [{\citenamefont {Huckestein}(1995)}]{Huckestein-Scaling-1995}%
  \BibitemOpen
  \bibfield  {author} {\bibinfo {author} {\bibfnamefont {B.}~\bibnamefont
  {Huckestein}},\ }\href {\doibase 10.1103/RevModPhys.67.357} {\bibfield
  {journal} {\bibinfo  {journal} {Rev. Mod. Phys.}\ }\textbf {\bibinfo {volume}
  {67}},\ \bibinfo {pages} {357} (\bibinfo {year} {1995})}\BibitemShut
  {NoStop}%
\bibitem [{\citenamefont {Huckestein}(1992)}]{Huckestein-1992}%
  \BibitemOpen
  \bibfield  {author} {\bibinfo {author} {\bibfnamefont {B.}~\bibnamefont
  {Huckestein}},\ }\href@noop {} {\bibfield  {journal} {\bibinfo  {journal}
  {Europhys. Lett.}\ }\textbf {\bibinfo {volume} {20}},\ \bibinfo {pages} {451}
  (\bibinfo {year} {1992})}\BibitemShut {NoStop}%
\bibitem [{\citenamefont {Huckestein}(1994)}]{Huckestein-1994}%
  \BibitemOpen
  \bibfield  {author} {\bibinfo {author} {\bibfnamefont {B.}~\bibnamefont
  {Huckestein}},\ }\href@noop {} {\bibfield  {journal} {\bibinfo  {journal}
  {Phys.Rev. Lett.}\ }\textbf {\bibinfo {volume} {72}},\ \bibinfo {pages}
  {1080} (\bibinfo {year} {1994})}\BibitemShut {NoStop}%
\bibitem [{\citenamefont {{Lee}}\ \emph {et~al.}(1993)\citenamefont {{Lee}},
  \citenamefont {{Wang}},\ and\ \citenamefont {{Kivelson}}}]{Lee-Quantum-1993}%
  \BibitemOpen
  \bibfield  {author} {\bibinfo {author} {\bibfnamefont {D.-H.}\ \bibnamefont
  {{Lee}}}, \bibinfo {author} {\bibfnamefont {Z.}~\bibnamefont {{Wang}}}, \
  and\ \bibinfo {author} {\bibfnamefont {S.}~\bibnamefont {{Kivelson}}},\
  }\href {\doibase 10.1103/PhysRevLett.70.4130} {\bibfield  {journal} {\bibinfo
   {journal} {\prl}\ }\textbf {\bibinfo {volume} {70}},\ \bibinfo {pages}
  {4130} (\bibinfo {year} {1993})}\BibitemShut {NoStop}%
\bibitem [{\citenamefont {{Ludwig}}\ \emph {et~al.}(1994)\citenamefont
  {{Ludwig}}, \citenamefont {{Fisher}}, \citenamefont {{Shankar}},\ and\
  \citenamefont {{Grinstein}}}]{Ludwig-Integer-1994}%
  \BibitemOpen
  \bibfield  {author} {\bibinfo {author} {\bibfnamefont {A.~W.~W.}\
  \bibnamefont {{Ludwig}}}, \bibinfo {author} {\bibfnamefont {M.~P.~A.}\
  \bibnamefont {{Fisher}}}, \bibinfo {author} {\bibfnamefont {R.}~\bibnamefont
  {{Shankar}}}, \ and\ \bibinfo {author} {\bibfnamefont {G.}~\bibnamefont
  {{Grinstein}}},\ }\href {\doibase 10.1103/PhysRevB.50.7526} {\bibfield
  {journal} {\bibinfo  {journal} {\prb}\ }\textbf {\bibinfo {volume} {50}},\
  \bibinfo {pages} {7526} (\bibinfo {year} {1994})}\BibitemShut {NoStop}%
\bibitem [{\citenamefont {{Yang}}\ and\ \citenamefont
  {{Bhatt}}(1996)}]{Bhatt-1996}%
  \BibitemOpen
  \bibfield  {author} {\bibinfo {author} {\bibfnamefont {K.}~\bibnamefont
  {{Yang}}}\ and\ \bibinfo {author} {\bibfnamefont {R.~N.}\ \bibnamefont
  {{Bhatt}}},\ }\href {\doibase 10.1103/PhysRevLett.76.1316} {\bibfield
  {journal} {\bibinfo  {journal} {Phys.Rev. Lett.}\ }\textbf {\bibinfo {volume}
  {76}},\ \bibinfo {pages} {1316} (\bibinfo {year} {1996})}\BibitemShut
  {NoStop}%
\bibitem [{\citenamefont {{de C.~Chamon}}\ \emph {et~al.}(1996)\citenamefont
  {{de C.~Chamon}}, \citenamefont {{Mudry}},\ and\ \citenamefont
  {{Wen}}}]{Chamon-Instability-1996}%
  \BibitemOpen
  \bibfield  {author} {\bibinfo {author} {\bibfnamefont {C.}~\bibnamefont {{de
  C.~Chamon}}}, \bibinfo {author} {\bibfnamefont {C.}~\bibnamefont {{Mudry}}},
  \ and\ \bibinfo {author} {\bibfnamefont {X.-G.}\ \bibnamefont {{Wen}}},\
  }\href {\doibase 10.1103/PhysRevB.53.R7638} {\bibfield  {journal} {\bibinfo
  {journal} {\prb}\ }\textbf {\bibinfo {volume} {53}},\ \bibinfo {pages} {7638}
  (\bibinfo {year} {1996})}\BibitemShut {NoStop}%
\bibitem [{\citenamefont {{Lee}}\ and\ \citenamefont
  {{Wang}}(1996)}]{Lee-Effects-1996}%
  \BibitemOpen
  \bibfield  {author} {\bibinfo {author} {\bibfnamefont {D.-H.}\ \bibnamefont
  {{Lee}}}\ and\ \bibinfo {author} {\bibfnamefont {Z.}~\bibnamefont {{Wang}}},\
  }\href {\doibase 10.1103/PhysRevLett.76.4014} {\bibfield  {journal} {\bibinfo
   {journal} {\prl}\ }\textbf {\bibinfo {volume} {76}},\ \bibinfo {pages}
  {4014} (\bibinfo {year} {1996})}\BibitemShut {NoStop}%
\bibitem [{\citenamefont {{Klesse}}\ and\ \citenamefont
  {{Metzler}}(1995)}]{Klesse-Universal-1995}%
  \BibitemOpen
  \bibfield  {author} {\bibinfo {author} {\bibfnamefont {R.}~\bibnamefont
  {{Klesse}}}\ and\ \bibinfo {author} {\bibfnamefont {M.}~\bibnamefont
  {{Metzler}}},\ }\href {\doibase 10.1209/0295-5075/32/3/007} {\bibfield
  {journal} {\bibinfo  {journal} {Europhys. Lett.}\ }\textbf {\bibinfo {volume}
  {32}},\ \bibinfo {pages} {229} (\bibinfo {year} {1995})}\BibitemShut
  {NoStop}%
\bibitem [{\citenamefont {Li}\ \emph {et~al.}(2005)\citenamefont {Li},
  \citenamefont {Cs\'athy}, \citenamefont {Tsui}, \citenamefont {Pfeiffer},\
  and\ \citenamefont {West}}]{Li-Scaling-2005}%
  \BibitemOpen
  \bibfield  {author} {\bibinfo {author} {\bibfnamefont {W.}~\bibnamefont
  {Li}}, \bibinfo {author} {\bibfnamefont {G.~A.}\ \bibnamefont {Cs\'athy}},
  \bibinfo {author} {\bibfnamefont {D.~C.}\ \bibnamefont {Tsui}}, \bibinfo
  {author} {\bibfnamefont {L.~N.}\ \bibnamefont {Pfeiffer}}, \ and\ \bibinfo
  {author} {\bibfnamefont {K.~W.}\ \bibnamefont {West}},\ }\href {\doibase
  10.1103/PhysRevLett.94.206807} {\bibfield  {journal} {\bibinfo  {journal}
  {Phys. Rev. Lett.}\ }\textbf {\bibinfo {volume} {94}},\ \bibinfo {pages}
  {206807} (\bibinfo {year} {2005})}\BibitemShut {NoStop}%
\bibitem [{\citenamefont {{Zirnbauer}}(1999)}]{Zirnbauer-Conformal-1999}%
  \BibitemOpen
  \bibfield  {author} {\bibinfo {author} {\bibfnamefont {M.~R.}\ \bibnamefont
  {{Zirnbauer}}},\ }\href@noop {} {\bibfield  {journal} {\bibinfo  {journal}
  {arXiv eprint}\ } (\bibinfo {year} {1999})},\ \Eprint
  {http://arxiv.org/abs/hep-th/9905054} {hep-th/9905054} \BibitemShut {NoStop}%
\bibitem [{\citenamefont {{Zirnbauer}}(1996)}]{Zirnbauer-Riemannian-1996}%
  \BibitemOpen
  \bibfield  {author} {\bibinfo {author} {\bibfnamefont {M.~R.}\ \bibnamefont
  {{Zirnbauer}}},\ }\href {\doibase 10.1063/1.531675} {\bibfield  {journal}
  {\bibinfo  {journal} {J. Math. Phys.}\ }\textbf {\bibinfo {volume} {37}},\
  \bibinfo {pages} {4986} (\bibinfo {year} {1996})}\BibitemShut {NoStop}%
\bibitem [{\citenamefont {Zirnbauer}(1994)}]{Zirnbauer-Towards-1994}%
  \BibitemOpen
  \bibfield  {author} {\bibinfo {author} {\bibfnamefont {M.~R.}\ \bibnamefont
  {Zirnbauer}},\ }\href@noop {} {\bibfield  {journal} {\bibinfo  {journal}
  {Annalen Phys.}\ }\textbf {\bibinfo {volume} {3}},\ \bibinfo {pages} {513}
  (\bibinfo {year} {1994})},\ \bibinfo {note} {[Erratum: Annalen Phys. {\bf 4},
  89 (1995)]}\BibitemShut {NoStop}%
%%CITATION = ANPYA,3,513;%%
\bibitem [{\citenamefont {{Galstyan}}\ and\ \citenamefont
  {{Raikh}}(1997)}]{Galstyan-Raikh-1997}%
  \BibitemOpen
  \bibfield  {author} {\bibinfo {author} {\bibfnamefont {A.~G.}\ \bibnamefont
  {{Galstyan}}}\ and\ \bibinfo {author} {\bibfnamefont {M.~E.}\ \bibnamefont
  {{Raikh}}},\ }\href@noop {} {\bibfield  {journal} {\bibinfo  {journal} {Phys.
  Rev. B}\ }\textbf {\bibinfo {volume} {56}},\ \bibinfo {pages} {1422}
  (\bibinfo {year} {1997})}\BibitemShut {NoStop}%
\bibitem [{\citenamefont {{Cain}}\ \emph {et~al.}(2003)\citenamefont {{Cain}},
  \citenamefont {{R\"{o}mer}},\ and\ \citenamefont
  {{Raikh}}}]{Cain-Romer-Raikh-2003}%
  \BibitemOpen
  \bibfield  {author} {\bibinfo {author} {\bibfnamefont {P.}~\bibnamefont
  {{Cain}}}, \bibinfo {author} {\bibfnamefont {R.}~\bibnamefont {{R\"{o}mer}}},
  \ and\ \bibinfo {author} {\bibfnamefont {M.~E.}\ \bibnamefont {{Raikh}}},\
  }\href {\doibase 10.1103/PhysRevB.56.1422} {\bibfield  {journal} {\bibinfo
  {journal} {Phys. Rev. B}\ }\textbf {\bibinfo {volume} {67}},\ \bibinfo
  {pages} {075307} (\bibinfo {year} {2003})}\BibitemShut {NoStop}%
\bibitem [{\citenamefont {{Mkhitaryan}}\ and\ \citenamefont
  {{Raikh}}(2009)}]{Mkhitaryan-Raikh-2009}%
  \BibitemOpen
  \bibfield  {author} {\bibinfo {author} {\bibfnamefont {V.~V.}\ \bibnamefont
  {{Mkhitaryan}}}\ and\ \bibinfo {author} {\bibfnamefont {M.~E.}\ \bibnamefont
  {{Raikh}}},\ }\href {\doibase 10.1103/PhysRevB.78.195409} {\bibfield
  {journal} {\bibinfo  {journal} {Phys. Rev. B}\ }\textbf {\bibinfo {volume}
  {79}},\ \bibinfo {pages} {125401} (\bibinfo {year} {2009})}\BibitemShut
  {NoStop}%
\bibitem [{\citenamefont {{Mkhitaryan}}\ \emph {et~al.}(2009)\citenamefont
  {{Mkhitaryan}}, \citenamefont {{Kagalovsky}},\ and\ \citenamefont
  {{Raikh}}}]{Mkhitaryan-Kagalovsky-Raikh-2009}%
  \BibitemOpen
  \bibfield  {author} {\bibinfo {author} {\bibfnamefont {V.~V.}\ \bibnamefont
  {{Mkhitaryan}}}, \bibinfo {author} {\bibfnamefont {V.}~\bibnamefont
  {{Kagalovsky}}}, \ and\ \bibinfo {author} {\bibfnamefont {M.~E.}\
  \bibnamefont {{Raikh}}},\ }\href {\doibase 10.1103/PhysRevLett.103.066801}
  {\bibfield  {journal} {\bibinfo  {journal} {Phys. Rev. Lett.}\ }\textbf
  {\bibinfo {volume} {103}},\ \bibinfo {pages} {066801} (\bibinfo {year}
  {2009})}\BibitemShut {NoStop}%
\bibitem [{\citenamefont {{Song}}\ and\ \citenamefont
  {{Prodan}}(2014)}]{Song-Prodan-2013}%
  \BibitemOpen
  \bibfield  {author} {\bibinfo {author} {\bibfnamefont {J.}~\bibnamefont
  {{Song}}}\ and\ \bibinfo {author} {\bibfnamefont {E.}~\bibnamefont
  {{Prodan}}},\ }\href {\doibase 10.1209/0295-5075/105/37001} {\bibfield
  {journal} {\bibinfo  {journal} {Euro.Phys. Lett.}\ }\textbf {\bibinfo
  {volume} {105}},\ \bibinfo {pages} {37001} (\bibinfo {year}
  {2014})}\BibitemShut {NoStop}%
\bibitem [{\citenamefont {{Bhaseen}}\ \emph {et~al.}(2000)\citenamefont
  {{Bhaseen}}, \citenamefont {{Kogan}}, \citenamefont {{Soloviev}},
  \citenamefont {{Taniguchi}},\ and\ \citenamefont
  {{Tsvelik}}}]{Bhaseen-Towards-2000}%
  \BibitemOpen
  \bibfield  {author} {\bibinfo {author} {\bibfnamefont {M.~J.}\ \bibnamefont
  {{Bhaseen}}}, \bibinfo {author} {\bibfnamefont {I.~I.}\ \bibnamefont
  {{Kogan}}}, \bibinfo {author} {\bibfnamefont {O.~A.}\ \bibnamefont
  {{Soloviev}}}, \bibinfo {author} {\bibfnamefont {N.}~\bibnamefont
  {{Taniguchi}}}, \ and\ \bibinfo {author} {\bibfnamefont {A.~M.}\ \bibnamefont
  {{Tsvelik}}},\ }\href {\doibase 10.1016/S0550-3213(00)00276-5} {\bibfield
  {journal} {\bibinfo  {journal} {Nucl. Phys. B}\ }\textbf {\bibinfo {volume}
  {580}},\ \bibinfo {pages} {688} (\bibinfo {year} {2000})}\BibitemShut
  {NoStop}%
\bibitem [{\citenamefont {{Tsvelik}}(2001)}]{Tsvelik-Wave-2001}%
  \BibitemOpen
  \bibfield  {author} {\bibinfo {author} {\bibfnamefont {A.~M.}\ \bibnamefont
  {{Tsvelik}}},\ }\href@noop {} {\bibfield  {journal} {\bibinfo  {journal}
  {arXiv eprint}\ } (\bibinfo {year} {2001})},\ \Eprint
  {http://arxiv.org/abs/cond-mat/0112008} {cond-mat/0112008} \BibitemShut
  {NoStop}%
\bibitem [{\citenamefont {LeClair}(2001)}]{Leclair2001}%
  \BibitemOpen
  \bibfield  {author} {\bibinfo {author} {\bibfnamefont {A.}~\bibnamefont
  {LeClair}},\ }\href {\doibase 10.1103/PhysRevB.64.045329} {\bibfield
  {journal} {\bibinfo  {journal} {Phys. Rev. B}\ }\textbf {\bibinfo {volume}
  {64}},\ \bibinfo {pages} {045329} (\bibinfo {year} {2001})}\BibitemShut
  {NoStop}%
\bibitem [{\citenamefont {{Kagalovsky}}\ \emph {et~al.}(1999)\citenamefont
  {{Kagalovsky}}, \citenamefont {{Horovitz}}, \citenamefont {{Avishai}},\ and\
  \citenamefont {{Chalker}}}]{Kagalovsky-Quantum-1999}%
  \BibitemOpen
  \bibfield  {author} {\bibinfo {author} {\bibfnamefont {V.}~\bibnamefont
  {{Kagalovsky}}}, \bibinfo {author} {\bibfnamefont {B.}~\bibnamefont
  {{Horovitz}}}, \bibinfo {author} {\bibfnamefont {Y.}~\bibnamefont
  {{Avishai}}}, \ and\ \bibinfo {author} {\bibfnamefont {J.~T.}\ \bibnamefont
  {{Chalker}}},\ }\href {\doibase 10.1103/PhysRevLett.82.3516} {\bibfield
  {journal} {\bibinfo  {journal} {\prl}\ }\textbf {\bibinfo {volume} {82}},\
  \bibinfo {pages} {3516} (\bibinfo {year} {1999})}\BibitemShut {NoStop}%
\bibitem [{\citenamefont {{Gruzberg}}\ \emph {et~al.}(1999)\citenamefont
  {{Gruzberg}}, \citenamefont {{Ludwig}},\ and\ \citenamefont
  {{Read}}}]{Gruzberg-Exact-1999}%
  \BibitemOpen
  \bibfield  {author} {\bibinfo {author} {\bibfnamefont {I.~A.}\ \bibnamefont
  {{Gruzberg}}}, \bibinfo {author} {\bibfnamefont {A.~W.~W.}\ \bibnamefont
  {{Ludwig}}}, \ and\ \bibinfo {author} {\bibfnamefont {N.}~\bibnamefont
  {{Read}}},\ }\href {\doibase 10.1103/PhysRevLett.82.4524} {\bibfield
  {journal} {\bibinfo  {journal} {\prl}\ }\textbf {\bibinfo {volume} {82}},\
  \bibinfo {pages} {4524} (\bibinfo {year} {1999})}\BibitemShut {NoStop}%
\bibitem [{\citenamefont {{Slevin}}\ and\ \citenamefont
  {{Ohtsuki}}(2009)}]{Slevin-Critical-2009}%
  \BibitemOpen
  \bibfield  {author} {\bibinfo {author} {\bibfnamefont {K.}~\bibnamefont
  {{Slevin}}}\ and\ \bibinfo {author} {\bibfnamefont {T.}~\bibnamefont
  {{Ohtsuki}}},\ }\href {\doibase 10.1103/PhysRevB.80.041304} {\bibfield
  {journal} {\bibinfo  {journal} {\prb}\ }\textbf {\bibinfo {volume} {80}},\
  \bibinfo {eid} {041304} (\bibinfo {year} {2009})}\BibitemShut {NoStop}%
\bibitem [{\citenamefont {{Amado}}\ \emph {et~al.}(2011)\citenamefont
  {{Amado}}, \citenamefont {{Malyshev}}, \citenamefont {{Sedrakyan}},\ and\
  \citenamefont {{Dom{\'{\i}}nguez-Adame}}}]{Amado-Numerical-2011}%
  \BibitemOpen
  \bibfield  {author} {\bibinfo {author} {\bibfnamefont {M.}~\bibnamefont
  {{Amado}}}, \bibinfo {author} {\bibfnamefont {A.~V.}\ \bibnamefont
  {{Malyshev}}}, \bibinfo {author} {\bibfnamefont {A.}~\bibnamefont
  {{Sedrakyan}}}, \ and\ \bibinfo {author} {\bibfnamefont {F.}~\bibnamefont
  {{Dom{\'{\i}}nguez-Adame}}},\ }\href {\doibase
  10.1103/PhysRevLett.107.066402} {\bibfield  {journal} {\bibinfo  {journal}
  {\prl}\ }\textbf {\bibinfo {volume} {107}},\ \bibinfo {eid} {066402}
  (\bibinfo {year} {2011})}\BibitemShut {NoStop}%
\bibitem [{\citenamefont {{Slevin}}\ and\ \citenamefont
  {{Ohtsuki}}(2012)}]{Slevin-Finite-2012}%
  \BibitemOpen
  \bibfield  {author} {\bibinfo {author} {\bibfnamefont {K.}~\bibnamefont
  {{Slevin}}}\ and\ \bibinfo {author} {\bibfnamefont {T.}~\bibnamefont
  {{Ohtsuki}}},\ }\href {\doibase 10.1142/S2010194512006162} {\bibfield
  {journal} {\bibinfo  {journal} {Int. J. Mod. Phys. Conf. Ser.}\ }\textbf
  {\bibinfo {volume} {11}},\ \bibinfo {pages} {60} (\bibinfo {year}
  {2012})}\BibitemShut {NoStop}%
\bibitem [{\citenamefont {{Obuse}}\ \emph {et~al.}(2012)\citenamefont
  {{Obuse}}, \citenamefont {{Gruzberg}},\ and\ \citenamefont
  {{Evers}}}]{Obuse-Finite-2012}%
  \BibitemOpen
  \bibfield  {author} {\bibinfo {author} {\bibfnamefont {H.}~\bibnamefont
  {{Obuse}}}, \bibinfo {author} {\bibfnamefont {I.~A.}\ \bibnamefont
  {{Gruzberg}}}, \ and\ \bibinfo {author} {\bibfnamefont {F.}~\bibnamefont
  {{Evers}}},\ }\href {\doibase 10.1103/PhysRevLett.109.206804} {\bibfield
  {journal} {\bibinfo  {journal} {\prl}\ }\textbf {\bibinfo {volume} {109}},\
  \bibinfo {eid} {206804} (\bibinfo {year} {2012})}\BibitemShut {NoStop}%
\bibitem [{\citenamefont {{Dahlhaus}}\ \emph {et~al.}(2011)\citenamefont
  {{Dahlhaus}}, \citenamefont {{Edge}}, \citenamefont {{Tworzyd{\l}o}},\ and\
  \citenamefont {{Beenakker}}}]{Dahlhaus-Quantum-2011}%
  \BibitemOpen
  \bibfield  {author} {\bibinfo {author} {\bibfnamefont {J.~P.}\ \bibnamefont
  {{Dahlhaus}}}, \bibinfo {author} {\bibfnamefont {J.~M.}\ \bibnamefont
  {{Edge}}}, \bibinfo {author} {\bibfnamefont {J.}~\bibnamefont
  {{Tworzyd{\l}o}}}, \ and\ \bibinfo {author} {\bibfnamefont {C.~W.~J.}\
  \bibnamefont {{Beenakker}}},\ }\href {\doibase 10.1103/PhysRevB.84.115133}
  {\bibfield  {journal} {\bibinfo  {journal} {\prb}\ }\textbf {\bibinfo
  {volume} {84}},\ \bibinfo {eid} {115133} (\bibinfo {year}
  {2011})}\BibitemShut {NoStop}%
\bibitem [{\citenamefont {{Nuding}}\ \emph {et~al.}(2015)\citenamefont
  {{Nuding}}, \citenamefont {{Kl{\"u}mper}},\ and\ \citenamefont
  {{Sedrakyan}}}]{Nuding-Localization-2015}%
  \BibitemOpen
  \bibfield  {author} {\bibinfo {author} {\bibfnamefont {W.}~\bibnamefont
  {{Nuding}}}, \bibinfo {author} {\bibfnamefont {A.}~\bibnamefont
  {{Kl{\"u}mper}}}, \ and\ \bibinfo {author} {\bibfnamefont {A.}~\bibnamefont
  {{Sedrakyan}}},\ }\href {\doibase 10.1103/PhysRevB.91.115107} {\bibfield
  {journal} {\bibinfo  {journal} {\prb}\ }\textbf {\bibinfo {volume} {91}},\
  \bibinfo {eid} {115107} (\bibinfo {year} {2015})}\BibitemShut {NoStop}%
\bibitem [{\citenamefont {Wei}\ \emph {et~al.}(1994)\citenamefont {Wei},
  \citenamefont {Engel},\ and\ \citenamefont {Tsui}}]{Wei-Current-1994}%
  \BibitemOpen
  \bibfield  {author} {\bibinfo {author} {\bibfnamefont {H.~P.}\ \bibnamefont
  {Wei}}, \bibinfo {author} {\bibfnamefont {L.~W.}\ \bibnamefont {Engel}}, \
  and\ \bibinfo {author} {\bibfnamefont {D.~C.}\ \bibnamefont {Tsui}},\ }\href
  {\doibase 10.1103/PhysRevB.50.14609} {\bibfield  {journal} {\bibinfo
  {journal} {Phys. Rev. B}\ }\textbf {\bibinfo {volume} {50}},\ \bibinfo
  {pages} {14609} (\bibinfo {year} {1994})}\BibitemShut {NoStop}%
\bibitem [{\citenamefont {Li}\ \emph {et~al.}(2009)\citenamefont {Li},
  \citenamefont {Vicente}, \citenamefont {Xia}, \citenamefont {Pan},
  \citenamefont {Tsui}, \citenamefont {Pfeiffer},\ and\ \citenamefont
  {West}}]{Li-Scaling-2009}%
  \BibitemOpen
  \bibfield  {author} {\bibinfo {author} {\bibfnamefont {W.}~\bibnamefont
  {Li}}, \bibinfo {author} {\bibfnamefont {C.~L.}\ \bibnamefont {Vicente}},
  \bibinfo {author} {\bibfnamefont {J.~S.}\ \bibnamefont {Xia}}, \bibinfo
  {author} {\bibfnamefont {W.}~\bibnamefont {Pan}}, \bibinfo {author}
  {\bibfnamefont {D.~C.}\ \bibnamefont {Tsui}}, \bibinfo {author}
  {\bibfnamefont {L.~N.}\ \bibnamefont {Pfeiffer}}, \ and\ \bibinfo {author}
  {\bibfnamefont {K.~W.}\ \bibnamefont {West}},\ }\href {\doibase
  10.1103/PhysRevLett.102.216801} {\bibfield  {journal} {\bibinfo  {journal}
  {Phys. Rev. Lett.}\ }\textbf {\bibinfo {volume} {102}},\ \bibinfo {pages}
  {216801} (\bibinfo {year} {2009})}\BibitemShut {NoStop}%
\bibitem [{\citenamefont {{Puschmann}}\ \emph {et~al.}(2019)\citenamefont
  {{Puschmann}}, \citenamefont {{Cain}}, \citenamefont {{Schreiber}},\ and\
  \citenamefont {{Vojta}}}]{Vojta-2019}%
  \BibitemOpen
  \bibfield  {author} {\bibinfo {author} {\bibfnamefont {M.}~\bibnamefont
  {{Puschmann}}}, \bibinfo {author} {\bibfnamefont {M.}~\bibnamefont {{Cain}}},
  \bibinfo {author} {\bibfnamefont {M.}~\bibnamefont {{Schreiber}}}, \ and\
  \bibinfo {author} {\bibfnamefont {T.}~\bibnamefont {{Vojta}}},\ }\href@noop
  {} {\bibfield  {journal} {\bibinfo  {journal} {Phys. Rev. B}\ }\textbf
  {\bibinfo {volume} {99}},\ \bibinfo {pages} {121301} (\bibinfo {year}
  {2019})}\BibitemShut {NoStop}%
\bibitem [{\citenamefont {{Zhu}}\ \emph {et~al.}(2019)\citenamefont {{Zhu}},
  \citenamefont {{Wu}}, \citenamefont {{Bhatt}},\ and\ \citenamefont
  {{Wan}}}]{Bhatt-2019}%
  \BibitemOpen
  \bibfield  {author} {\bibinfo {author} {\bibfnamefont {Q.}~\bibnamefont
  {{Zhu}}}, \bibinfo {author} {\bibfnamefont {P.}~\bibnamefont {{Wu}}},
  \bibinfo {author} {\bibfnamefont {R.}~\bibnamefont {{Bhatt}}}, \ and\
  \bibinfo {author} {\bibfnamefont {X.}~\bibnamefont {{Wan}}},\ }\href@noop {}
  {\bibfield  {journal} {\bibinfo  {journal} {Phys. Rev. B}\ }\textbf {\bibinfo
  {volume} {99}},\ \bibinfo {pages} {024205} (\bibinfo {year}
  {2019})}\BibitemShut {NoStop}%
\bibitem [{\citenamefont {Polyakov}\ and\ \citenamefont
  {Shklovskii}(1993)}]{Polyakov-Shklovskii-1993}%
  \BibitemOpen
  \bibfield  {author} {\bibinfo {author} {\bibfnamefont {D.~G.}\ \bibnamefont
  {Polyakov}}\ and\ \bibinfo {author} {\bibfnamefont {B.~I.}\ \bibnamefont
  {Shklovskii}},\ }\href@noop {} {\bibfield  {journal} {\bibinfo  {journal}
  {Phys. Rev. Lett.}\ }\textbf {\bibinfo {volume} {70}},\ \bibinfo {pages}
  {3796} (\bibinfo {year} {1993})}\BibitemShut {NoStop}%
\bibitem [{\citenamefont {Pruisken}\ and\ \citenamefont
  {Baranov}(1995)}]{Pruisken-Baranov-1995}%
  \BibitemOpen
  \bibfield  {author} {\bibinfo {author} {\bibfnamefont {A.~M.~M.}\
  \bibnamefont {Pruisken}}\ and\ \bibinfo {author} {\bibfnamefont {M.~A.}\
  \bibnamefont {Baranov}},\ }\href@noop {} {\bibfield  {journal} {\bibinfo
  {journal} {Europhysics Letters}\ }\textbf {\bibinfo {volume} {31}},\ \bibinfo
  {pages} {543} (\bibinfo {year} {1995})}\BibitemShut {NoStop}%
\bibitem [{\citenamefont {Pruisken}\ and\ \citenamefont
  {Burmistrov}(2008)}]{Pruisken-Burmistrov-1995}%
  \BibitemOpen
  \bibfield  {author} {\bibinfo {author} {\bibfnamefont {A.~M.~M.}\
  \bibnamefont {Pruisken}}\ and\ \bibinfo {author} {\bibfnamefont {I.~S.}\
  \bibnamefont {Burmistrov}},\ }\href@noop {} {\bibfield  {journal} {\bibinfo
  {journal} {J. Exp. Theor.Phys. Lett.}\ ,\ \bibinfo {pages} {220}} (\bibinfo
  {year} {2008})}\BibitemShut {NoStop}%
\bibitem [{\citenamefont {Wang}\ \emph {et~al.}(2000)\citenamefont {Wang},
  \citenamefont {Fisher}, \citenamefont {Girvin},\ and\ \citenamefont
  {Chalker}}]{Girvin-Chalker-2000}%
  \BibitemOpen
  \bibfield  {author} {\bibinfo {author} {\bibfnamefont {Z.}~\bibnamefont
  {Wang}}, \bibinfo {author} {\bibfnamefont {M.~P.~A.}\ \bibnamefont {Fisher}},
  \bibinfo {author} {\bibfnamefont {S.~M.}\ \bibnamefont {Girvin}}, \ and\
  \bibinfo {author} {\bibfnamefont {J.~T.}\ \bibnamefont {Chalker}},\
  }\href@noop {} {\bibfield  {journal} {\bibinfo  {journal} {Phys. Rev. B}\
  }\textbf {\bibinfo {volume} {61}},\ \bibinfo {pages} {8326} (\bibinfo {year}
  {2000})}\BibitemShut {NoStop}%
\bibitem [{\citenamefont {Burmistrov}\ \emph {et~al.}(2011)\citenamefont
  {Burmistrov}, \citenamefont {Bera}, \citenamefont {Evers}, \citenamefont
  {Gornyi},\ and\ \citenamefont {Mirlin}}]{Burmistrov-2011}%
  \BibitemOpen
  \bibfield  {author} {\bibinfo {author} {\bibfnamefont {I.~S.}\ \bibnamefont
  {Burmistrov}}, \bibinfo {author} {\bibfnamefont {S.}~\bibnamefont {Bera}},
  \bibinfo {author} {\bibfnamefont {F.}~\bibnamefont {Evers}}, \bibinfo
  {author} {\bibfnamefont {I.~V.}\ \bibnamefont {Gornyi}}, \ and\ \bibinfo
  {author} {\bibfnamefont {A.~D.}\ \bibnamefont {Mirlin}},\ }\href@noop {}
  {\bibfield  {journal} {\bibinfo  {journal} {Annals of Physics}\ }\textbf
  {\bibinfo {volume} {326}},\ \bibinfo {pages} {1457} (\bibinfo {year}
  {2011})}\BibitemShut {NoStop}%
\bibitem [{\citenamefont {{Gruzberg}}\ \emph {et~al.}(2017)\citenamefont
  {{Gruzberg}}, \citenamefont {{Nuding}}, \citenamefont {{Kluemper}},\ and\
  \citenamefont {{Sedrakyan}}}]{GruzbergNudingKluemperSedrakyan2017}%
  \BibitemOpen
  \bibfield  {author} {\bibinfo {author} {\bibfnamefont {I.}~\bibnamefont
  {{Gruzberg}}}, \bibinfo {author} {\bibfnamefont {W.}~\bibnamefont
  {{Nuding}}}, \bibinfo {author} {\bibfnamefont {A.}~\bibnamefont
  {{Kluemper}}}, \ and\ \bibinfo {author} {\bibfnamefont {A.}~\bibnamefont
  {{Sedrakyan}}},\ }\href {\doibase 10.1103/PhysRevB.95.125414} {\bibfield
  {journal} {\bibinfo  {journal} {Phys. Rev. B}\ }\textbf {\bibinfo {volume}
  {95}},\ \bibinfo {pages} {125414} (\bibinfo {year} {2017})}\BibitemShut
  {NoStop}%
\bibitem [{\citenamefont {Ambj\"{o}rn}\ and\ \citenamefont
  {Sedrakyan}(2015)}]{Ambjorn-Sedrakyan-2013}%
  \BibitemOpen
  \bibfield  {author} {\bibinfo {author} {\bibfnamefont {J.}~\bibnamefont
  {Ambj\"{o}rn}}\ and\ \bibinfo {author} {\bibfnamefont {A.}~\bibnamefont
  {Sedrakyan}},\ }\href {\doibase 10.1016/j.nuclphysb.2013.06.017} {\bibfield
  {journal} {\bibinfo  {journal} {Nucl. Phys. B}\ }\textbf {\bibinfo {volume}
  {874 [PM]}},\ \bibinfo {pages} {877–888} (\bibinfo {year}
  {2015})}\BibitemShut {NoStop}%
\bibitem [{\citenamefont {Ambj\"{o}rn}\ \emph {et~al.}(2015)\citenamefont
  {Ambj\"{o}rn}, \citenamefont {Sh.},\ and\ \citenamefont
  {Sedrakyan}}]{Ambjorn-Sedrakyan-2015}%
  \BibitemOpen
  \bibfield  {author} {\bibinfo {author} {\bibfnamefont {J.}~\bibnamefont
  {Ambj\"{o}rn}}, \bibinfo {author} {\bibfnamefont {K.}~\bibnamefont {Sh.}}, \
  and\ \bibinfo {author} {\bibfnamefont {A.}~\bibnamefont {Sedrakyan}},\ }\href
  {\doibase 10.1103/PhysRevD.92.026002} {\bibfield  {journal} {\bibinfo
  {journal} {Phys. Rev. D}\ }\textbf {\bibinfo {volume} {92}},\ \bibinfo
  {pages} {026002} (\bibinfo {year} {2015})}\BibitemShut {NoStop}%
\bibitem [{\citenamefont {{Can}}\ \emph {et~al.}(2014)\citenamefont {{Can}},
  \citenamefont {{Laskin}},\ and\ \citenamefont
  {{Wiegmann}}}]{Can-Laskin-Wiegmann-2014}%
  \BibitemOpen
  \bibfield  {author} {\bibinfo {author} {\bibfnamefont {T.}~\bibnamefont
  {{Can}}}, \bibinfo {author} {\bibfnamefont {M.}~\bibnamefont {{Laskin}}}, \
  and\ \bibinfo {author} {\bibfnamefont {P.}~\bibnamefont {{Wiegmann}}},\
  }\href {\doibase 10.1103/PhysRevLett.113.046803} {\bibfield  {journal}
  {\bibinfo  {journal} {Phys. Rev. Lett.}\ }\textbf {\bibinfo {volume} {113}},\
  \bibinfo {pages} {046803} (\bibinfo {year} {2014})}\BibitemShut {NoStop}%
\bibitem [{\citenamefont {{Can}}\ \emph {et~al.}(2015)\citenamefont {{Can}},
  \citenamefont {{Laskin}},\ and\ \citenamefont
  {{Wiegmann}}}]{Can-Laskin-Wiegmann-2015}%
  \BibitemOpen
  \bibfield  {author} {\bibinfo {author} {\bibfnamefont {T.}~\bibnamefont
  {{Can}}}, \bibinfo {author} {\bibfnamefont {M.}~\bibnamefont {{Laskin}}}, \
  and\ \bibinfo {author} {\bibfnamefont {P.}~\bibnamefont {{Wiegmann}}},\
  }\href {\doibase 10.1016/j.aop.2015.02.013} {\bibfield  {journal} {\bibinfo
  {journal} {Ann. Phys.}\ }\textbf {\bibinfo {volume} {362}},\ \bibinfo {pages}
  {752} (\bibinfo {year} {2015})}\BibitemShut {NoStop}%
\bibitem [{\citenamefont {{Laskin}}\ \emph {et~al.}(2015)\citenamefont
  {{Laskin}}, \citenamefont {{Can}},\ and\ \citenamefont
  {{Wiegmann}}}]{Laskin-Can-Wiegmann-2015}%
  \BibitemOpen
  \bibfield  {author} {\bibinfo {author} {\bibfnamefont {M.}~\bibnamefont
  {{Laskin}}}, \bibinfo {author} {\bibfnamefont {T.}~\bibnamefont {{Can}}}, \
  and\ \bibinfo {author} {\bibfnamefont {P.}~\bibnamefont {{Wiegmann}}},\
  }\href {\doibase 10.1103/PhysRevB.92.235141} {\bibfield  {journal} {\bibinfo
  {journal} {Phys. Rev. B}\ }\textbf {\bibinfo {volume} {92}},\ \bibinfo
  {pages} {235141} (\bibinfo {year} {2015})}\BibitemShut {NoStop}%
\bibitem [{\citenamefont {Harris}(1974)}]{Harris-1974}%
  \BibitemOpen
  \bibfield  {author} {\bibinfo {author} {\bibfnamefont {A.~B.}\ \bibnamefont
  {Harris}},\ }\href@noop {} {\bibfield  {journal} {\bibinfo  {journal} {J.
  Phys. C}\ }\textbf {\bibinfo {volume} {7}},\ \bibinfo {pages} {1671}
  (\bibinfo {year} {1974})}\BibitemShut {NoStop}%
\bibitem [{\citenamefont {Chayes}\ \emph {et~al.}(1986)\citenamefont {Chayes},
  \citenamefont {Chayes}, \citenamefont {Fisher},\ and\ \citenamefont
  {Spencer}}]{Chayes-Chayes-1986}%
  \BibitemOpen
  \bibfield  {author} {\bibinfo {author} {\bibfnamefont {J.~T.}\ \bibnamefont
  {Chayes}}, \bibinfo {author} {\bibfnamefont {L.}~\bibnamefont {Chayes}},
  \bibinfo {author} {\bibfnamefont {D.~S.}\ \bibnamefont {Fisher}}, \ and\
  \bibinfo {author} {\bibfnamefont {T.}~\bibnamefont {Spencer}},\ }\href@noop
  {} {\bibfield  {journal} {\bibinfo  {journal} {Phys. Rev. Lett.}\ }\textbf
  {\bibinfo {volume} {57}},\ \bibinfo {pages} {2999} (\bibinfo {year}
  {1986})}\BibitemShut {NoStop}%
\bibitem [{\citenamefont {Vojta}(2013)}]{Vojta-2013}%
  \BibitemOpen
  \bibfield  {author} {\bibinfo {author} {\bibfnamefont {T.}~\bibnamefont
  {Vojta}},\ }\href {\doibase 10.1063/1.4818403} {\bibfield  {journal}
  {\bibinfo  {journal} {Lectures on the Physics of Strongly Correlated Systems
  XVII AIP Conf. Proc.}\ }\textbf {\bibinfo {volume} {1550}},\ \bibinfo {pages}
  {188} (\bibinfo {year} {2013})}\BibitemShut {NoStop}%
\bibitem [{\citenamefont {MacKinnon}\ and\ \citenamefont
  {Kramer}(1981)}]{MacKinnon-One-Parameter-1981}%
  \BibitemOpen
  \bibfield  {author} {\bibinfo {author} {\bibfnamefont {A.}~\bibnamefont
  {MacKinnon}}\ and\ \bibinfo {author} {\bibfnamefont {B.}~\bibnamefont
  {Kramer}},\ }\href {\doibase 10.1103/PhysRevLett.47.1546} {\bibfield
  {journal} {\bibinfo  {journal} {Phys. Rev. Lett.}\ }\textbf {\bibinfo
  {volume} {47}},\ \bibinfo {pages} {1546} (\bibinfo {year}
  {1981})}\BibitemShut {NoStop}%
\bibitem [{\citenamefont {{MacKinnon}}\ and\ \citenamefont
  {{Kramer}}(1983)}]{MacKinnon-The-scaling-1983}%
  \BibitemOpen
  \bibfield  {author} {\bibinfo {author} {\bibfnamefont {A.}~\bibnamefont
  {{MacKinnon}}}\ and\ \bibinfo {author} {\bibfnamefont {B.}~\bibnamefont
  {{Kramer}}},\ }\href {\doibase 10.1007/BF01578242} {\bibfield  {journal}
  {\bibinfo  {journal} {Z. Phys. B}\ }\textbf {\bibinfo {volume} {53}},\
  \bibinfo {pages} {1} (\bibinfo {year} {1983})}\BibitemShut {NoStop}%
\bibitem [{\citenamefont {Oseledec}(1968)}]{Oseledec-A-Multiplicative-1968}%
  \BibitemOpen
  \bibfield  {author} {\bibinfo {author} {\bibfnamefont {V.~I.}\ \bibnamefont
  {Oseledec}},\ }\href@noop {} {\bibfield  {journal} {\bibinfo  {journal}
  {Trudy Moskov. Mat. Ob\v s\v c.}\ }\textbf {\bibinfo {volume} {19}},\
  \bibinfo {pages} {179} (\bibinfo {year} {1968})}\BibitemShut {NoStop}%
\bibitem [{\citenamefont {Tutubalin}(1965)}]{Tutubalin-On-Limit-1965}%
  \BibitemOpen
  \bibfield  {author} {\bibinfo {author} {\bibfnamefont {V.}~\bibnamefont
  {Tutubalin}},\ }\href {\doibase 10.1137/1110002} {\bibfield  {journal}
  {\bibinfo  {journal} {Theory Probab. Appl.}\ }\textbf {\bibinfo {volume}
  {10}},\ \bibinfo {pages} {15} (\bibinfo {year} {1965})}\BibitemShut {NoStop}%
\bibitem [{\citenamefont {{Obuse}}\ \emph {et~al.}(2010)\citenamefont
  {{Obuse}}, \citenamefont {{Subramaniam}}, \citenamefont {{Furusaki}},
  \citenamefont {{Gruzberg}},\ and\ \citenamefont
  {{Ludwig}}}]{Obuse-Conformal-2010}%
  \BibitemOpen
  \bibfield  {author} {\bibinfo {author} {\bibfnamefont {H.}~\bibnamefont
  {{Obuse}}}, \bibinfo {author} {\bibfnamefont {A.~R.}\ \bibnamefont
  {{Subramaniam}}}, \bibinfo {author} {\bibfnamefont {A.}~\bibnamefont
  {{Furusaki}}}, \bibinfo {author} {\bibfnamefont {I.~A.}\ \bibnamefont
  {{Gruzberg}}}, \ and\ \bibinfo {author} {\bibfnamefont {A.~W.~W.}\
  \bibnamefont {{Ludwig}}},\ }\href {\doibase 10.1103/PhysRevB.82.035309}
  {\bibfield  {journal} {\bibinfo  {journal} {\prb}\ }\textbf {\bibinfo
  {volume} {82}},\ \bibinfo {eid} {035309} (\bibinfo {year}
  {2010})}\BibitemShut {NoStop}%
\bibitem [{\citenamefont {Akaike}(1974)}]{Akaike-A-new-1974}%
  \BibitemOpen
  \bibfield  {author} {\bibinfo {author} {\bibfnamefont {H.}~\bibnamefont
  {Akaike}},\ }\href {\doibase 10.1109/TAC.1974.1100705} {\bibfield  {journal}
  {\bibinfo  {journal} {IEEE Trans. Automat. Control}\ }\textbf {\bibinfo
  {volume} {19}},\ \bibinfo {pages} {716} (\bibinfo {year} {1974})}\BibitemShut
  {NoStop}%
\end{thebibliography}%
\bibliographystyle{apsrev4-1}

%\end{document}

%\appendix
\section*{SUPPLEMENTAL MATERIAL}
\setcounter{equation}{0}
\subsection*{The fitting procedure}

As is standard in the transfer matrix method, we want to numerically estimate
the Lyapunov exponent $\gamma$ defined as the smallest positive eigenvalue of
\bea 
\label{LE} 
\frac{1}{2L} \log[T_L T_L^\dagger]. 
\eea in the limit as $L
\to \infty$. 
This quantity is self-averaging, and for finite $L$ its
distribution is basically Gaussian. This is illustrated for a particular set
of parameters $M, X, p$ in Fig.~\ref{fig:histogram2}.
\begin{figure}[h]
	\centering
	\includegraphics{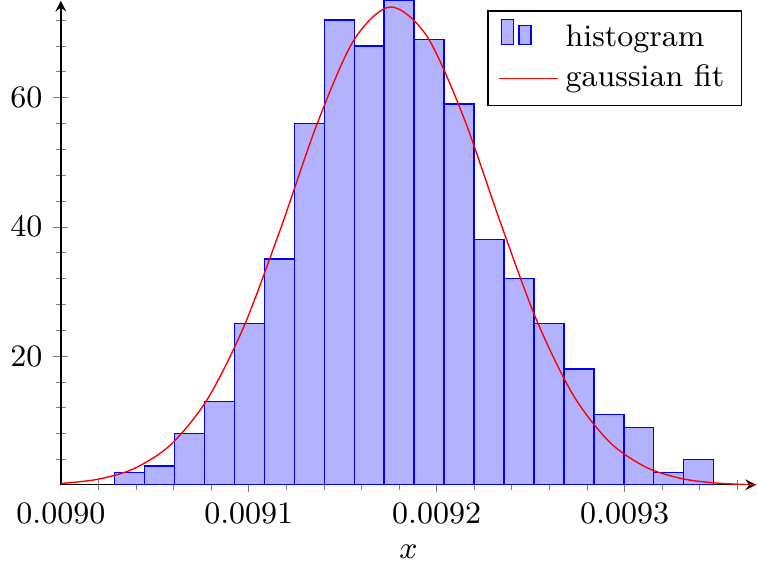}
	\caption {Distribution of Lyapunov exponents in {the
            ensemble of 624 realizations for chain length $L=5\,000\,000$,
            block size $M=100$, parameter $x=0.04$, and disorder parameter $p
            = 1/4$.}}
	\label{fig:histogram2}
\end{figure}

We numerically calculated $\gamma$ for
various combinations of the parameter $x$ and the lattice width $M$. The
results are shown in Fig.~\ref{fig:FitLines333}.

\begin{figure}[h]
	\centering
	\includegraphics[width=\columnwidth, angle=0]{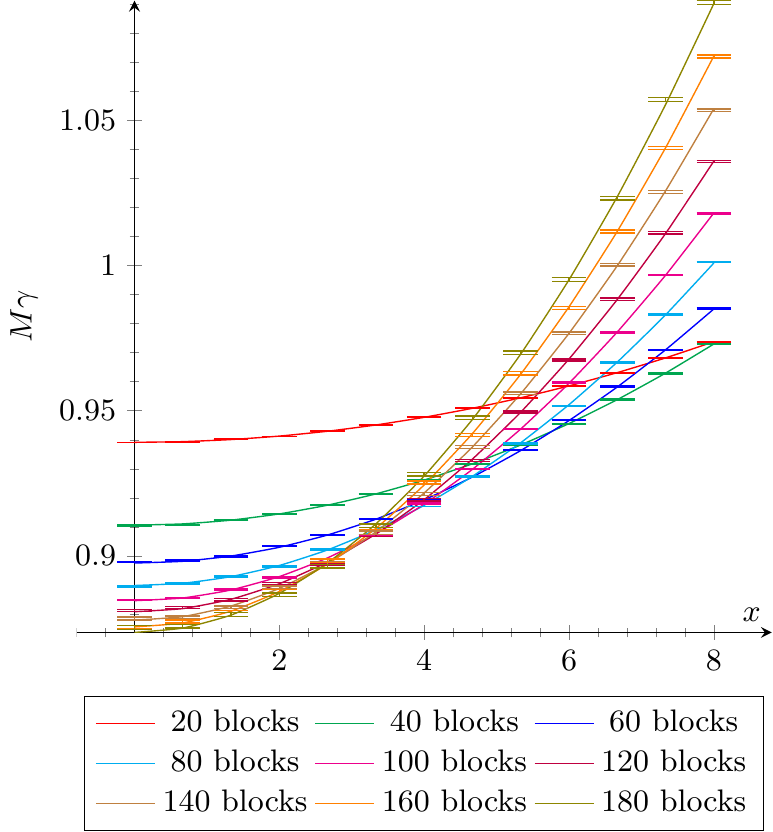}
    \caption 
    { 
    	Plot of the logarithm of the smallest eigenvalue of the transfer matrix
		times $M$ ($=$ number of blocks) depending on the distance $x$ from the
		critical point for the model with disorder parameter $p=1/4$. The $x$-values
		divide the interval $[0,0.08]$ into 12 equal parts.  The data points are
		given by the average of the ensemble belonging to the corresponding values
		for $x$ and $M$. All considered values for $M$ are listed in the legend. The
		product length is $L=5\,000\,000$. The error bars are obtained from the
		standard deviation of this ensemble. The curves are obtained by plotting the
		fit function for the relevant values of $M$ in the regime $x=0$ to $x=0.08$.
		The ensembles vary between 208 and 1096 eigenvalues. Details on the ensemble
		sizes can be found in appendix \ref{ensembleStats} 
	} \label{fig:FitLines333}
\end{figure}

It is clearly seen that the lines corresponding to different values of $M$ do
not intersect at the critical value $x = 0$. In fact, they do not intersect
at a single point at all. Therefore, any attempt at trying to use a
single-parameter scaling  to collapse the data is doomed to fail. The reason
for this is that the critical point of the CC model is not the same as the
fixed point. They differ by the presence of irrelevant variables that decay
as we increase the system width. For the CC model specifically, the leading
irrelevant variable has the scaling exponent $y < 0$ which is rather small in
magnitude. This causes strong correction to scaling even at the critical
point. This is a known feature of the CC model that has been stressed by
Slevin and Ohtsuki in Ref.~[\onlinecite{Slevin-Critical-2009}]. They
emphasized that it is crucial to include irrelevant scaling variables as
arguments of the fitting functions used in the scaling analysis of the data.
This procedure leads to much more reliable results, but cannot be visualized
as a simple scaling collapse of the numerical data, as in the case of a
single-variable scaling. Inclusion of irrelevant variables in the scaling
analysis has become a standard procedure in the numerical studies of network
models, and here we follow the same procedure.

Thus, we fit the scaling behavior of the Lyapunov exponent $\gamma$ near
the critical point to the following expression:
\begin{equation} \label{ren_equ}
	\gamma\cdot M=\Gamma(M^{1/\nu}u_0,M^y\,u_1) ,
\end{equation}
Here we have taken into account the relevant field with exponent $\nu$ and
the leading irrelevant field with exponent $y$. $M$ is the number of $2
\times 2$ blocks in the transfer matrices ($=$ half the number of horizontal
channels of the lattice), $u_0=u_0(x)$ is the relevant field and $u_1=u_1(x)$
the leading irrelevant field. It is known that the relevant field vanishes at
the critical point, and that $y<0$.

Regarding the two-variable fit, on the left hand side of Eq.~\eqref{ren_equ} we
use the numerical results for the eigenvalues of $T_L$, where we are
particularly interested in the eigenvalue closest to 1. The right hand side of
\eqref{ren_equ} is expanded in a series in $x$ and powers of $M$, and the
expansion coefficients are obtained from a fit. Some coefficients in this
expansion vanish due to a symmetry argument. \cite{Slevin-Critical-2009} If $x$
is replaced by $-x$ we see from \eqref{rt} that $t$ turns into $r$ and vice
versa. Due to the periodic boundary conditions the lattice is unchanged.
Therefore the left hand side of \eqref{ren_equ} is invariant under the sign
change of $x$. Hence the right hand side must be even in $x$. That renders
$u_0(x)$ and $u_1(x)$ either even or odd in $x$. For the Chalker Coddington
network the critical point is at $x=0$. 
{This lets us choose $u_0(x)$ odd and $u_1(x)$ even.} 
The fit now should
use as few coefficients as possible while reproducing the data as closely as
possible.

The scaling function $\Gamma$ in the right side of \eqref{ren_equ} is
expanded in the fields $u_0$ and $u_1$ yielding
\begin{equation}
\label{expansin_in_fields}
\begin{split}
	\Gamma(&u_0(x)M^{1/\nu},u_1(x)M^y)= \Gamma_c+ \Gamma_{01} u_1M^y
	 +\Gamma_{20}u_0^2M^{2/\nu}\\	& + \Gamma_{02}u_1^2M^{2y}
	 +\Gamma_{21}u_0^2u_1M^{2/\nu}M^y +\Gamma_{03}u_1^3 M^{3y} \\
	& +\Gamma_{40}u_0^4M^{4/\nu}+\Gamma_{22}u_0^2 M^{2/\nu}u_1^2 M^{2y} + \Gamma_{04}u_1^4M^{4y}+\dots
\end{split}
\end{equation}
We further expand $u_0$ and $u_1$ in powers of $x$ as was done, for example,
in Refs. [\onlinecite{Slevin-Critical-2009},
\onlinecite{Amado-Numerical-2011}]:
\begin{equation}
\label{fields_expanded}
 u_0(x)=x+\sum_{k=1}^\infty a_{2k+1}x^{2k+1} \quad \text{and} \quad
 u_1(x)=1+\sum_{k=1}^\infty b_{2k}x^{2k} .
\end{equation}
In Eq. \eqref{expansin_in_fields} we retained only terms that are even in
$x$. Because of the ambiguity in the overall scaling of the fields, the
leading coefficient in Eq. \eqref{fields_expanded} can be chosen to be 1.

The first term in the expansion (\ref{expansin_in_fields}), $\Gamma_c$
represents the asymptotic value of the universal critical amplitude ratio
$\Gamma$ in the infinite system. Theoretical arguments based on conformal
invariance relate $\Gamma$ to the multifractal exponent $\alpha_0$:
\begin{align}
\Gamma_c = \pi (\alpha_0 - 2),
\end{align}
see, for example, Ref. [\onlinecite{Obuse-Conformal-2010}].

\subsection*{Weights and Errors}\label{weight_error}

The left hand side of Eq.~\eqref{ren_equ} is determined by the results of
numerical simulations of the random network model. Following
Ref.~[\onlinecite{Amado-Numerical-2011}] we have produced large ensembles of the
Lyapunov exponent $\gamma$ for a variety of choices for the probability index
$p$ by simulating many disorder realizations for many combinations of $x$ and
$M$. We calculated disorder realizations for any combination of $M=20, 40, 60,
80, 100, 120, 140, 160, 180, 200$ and $x=0.08/12\cdot [0, 1, 2, 3, 4, 5, 6, 7,
8, 9, 10, 11, 12]$ for fixed $L=5\,000\,000$. Details on the ensemble sizes can
be found in appendix \ref{ensembleStats}. Our goal is to check whether the
central limit theorem \cite{Tutubalin-On-Limit-1965} also works in the case of
randomness of the network or not. Fig.~\ref{fig:histogram2} shows the
distribution of the Lyapunov exponent for $p=1/4$, $M=100$ and $x=0.04$ being
nicely described by a Gaussian which demonstrates the validity of the central
limit theorem.

In the fitting procedure, the weight of each such $\gamma$ is given by the
reciprocal of the {variance} of the corresponding ensemble. So all $\gamma$
from the same $(x,M)$ ensemble enter the fit with the same weight. On the
right hand side of Eq. \eqref{ren_equ} the fitting formula
\eqref{expansin_in_fields} depending on $x$
and $M$ is used. The coefficients of the expansion and the critical exponents
are the fitting coefficients.

The fits are performed in several steps. First a weighted nonlinear least
square fit based on a trust region algorithm with specified regions for each
parameter is applied. The resulting parameters are used in a further weighted
nonlinear least square fit based on a trust region algorithm. Here no limits
are imposed on the fit parameters.  The last step is repeated until the
resulting parameters stop changing.

\subsection*{Evaluation of fits}

There are several methods for the evaluation of the fit results.

The \emph{${\chi^2}$-test} with $\chi^2$ given by
\begin{equation}
	\chi^2=\sum_i \frac{(y_i-f_i)^2}{\sigma_i^2}
\end{equation}
where $f_i$ is the value obtained by the fit function and $y_i$ is the
measured value.  The parameters $\sigma_i$ are the standard deviations of the
ensemble $i$ with values for $(x_i,M_i)$.  Our fit contains a large ensemble
of data points for each $(x_i,M_i)$ coordinate. Hence $\chi^2=0$ is not
possible, in fact it will be large due to the huge number of data
points. Therefore, we consider the ratio $\chi^2$/\emph{degrees of freedom}
with expectation value 1 in case of an ideal fit.  The \emph{degrees of
  freedom} equals the number of data points in the fit minus the number of fit
parameters.

Deviations from 1 are evaluated by use of the cumulative probability
$P(\tilde\chi^2 < \chi^2)$ which is the probability of observing a sample
statistic with a smaller $\chi^2$ value than in our fit. A small value of $P$,
and hence a large value of the complement $Q:=1-P$ is indicative for a good
fit. Yet, values of $P$ lower than $1/2$ would indicate problems in the
estimation of error bars of the individual data points.

Another criterion uses the width of the \emph{confidence intervals} which
quantifies the quality of the prediction for a single parameter. We use 95\%
confidence intervals meaning that for repeated independent generations of the
data and subsequent data analysis, the resulting confidence intervals contain
the true parameter values in 95\% of the cases.

A very sensitive criterion is the \emph{Akaike information criterion} (AIC)
\cite{Akaike-A-new-1974} which allows to select between model fit functions.
Suppose, we have $l$ models with AIC$_1$, \dots AIC$_l$. The model with the
smallest AIC is the favorite one: The relative probability of
  model $j$ compared to the model with minimum AIC$_{min}$ is
\begin{align}
\exp \frac{\text{AIC}_{min}-\text{AIC}_j}{2},
\end{align}
which is always smaller than one.

The last criterion we present is the sum of \emph{residuals} which is given by	
$\mathit{res}=\sum_j \mathit{res}_j, \; \mathit{res}_j=y_j-f_j$. The condition
is that $\mathit{res}$ be small compared to the number of degrees of freedom. 

\onecolumngrid
%\newpage
\appendix
%\begin{widetext}

\section{Tables of ensemble statistics} \label{ensembleStats}
{In this appendix  we present the statistics of our data sets.}
For each $p$ we present a table showing the number of Lyapunov exponents 
for each $(x,M)$ pair. As explained in the introduction, $p$ is the probability 
for enforced horizontal respectively vertical transition in the network and $1-2p$ is 
the probability for regular scattering. \\

\noindent
$\bm{p=0}$ \textbf{(Classical Chalker Coddington Lattice \cite{Nuding-Localization-2015})} \\
\begin{tabular}{|r| *{13}{c} |c|}
	\hline
	\bm{$M$} & \multicolumn{13}{c}{\bm{$x$}} \vline & \bm{$L$} \\ 
	&	0	&	0.0067	&	0.0133	&	0.0200	&	0.0267	&	0.0333	&	0.0400	&	0.04667	&	0.0533	&	0.0600	&	0.0667	&	0.0733	&	0.0800	&	\\
	\hline
	20	&	208	&	208	&	208	&	208	&	208	&	208	&	208	&	208	&	208	&	208	&	208	&	208	&	208	&	5000000	\\
	40	&	152	&	152	&	304	&	152	&	152	&	0	&	152	&	152	&	152	&	152	&	152	&	152	&	152	&	5000000	\\
	60	&	208	&	208	&	208	&	208	&	208	&	208	&	208	&	208	&	208	&	208	&	208	&	208	&	208	&	5000000	\\
	80	&	208	&	208	&	208	&	208	&	208	&	208	&	208	&	208	&	208	&	208	&	208	&	208	&	208	&	5000000	\\
	100	&	208	&	200	&	200	&	200	&	200	&	200	&	200	&	200	&	200	&	200	&	208	&	200	&	200	&	5000000	\\
	120	&	150	&	150	&	296	&	146	&	148	&	0	&	150	&	152	&	150	&	150	&	152	&	150	&	150	&	5000000	\\
	140	&	208	&	208	&	208	&	208	&	208	&	208	&	208	&	208	&	208	&	208	&	208	&	208	&	208	&	5000000	\\
	160	&	208	&	176	&	144	&	144	&	176	&	112	&	144	&	192	&	192	&	144	&	128	&	144	&	128	&	5000000	\\
	180	&	208	&	208	&	208	&	208	&	208	&	208	&	208	&	208	&	208	&	208	&	208	&	208	&	208	&	5000000	\\
	\hline
\end{tabular} \\

\vspace{2ex}
\noindent
\bm{$p=0.1$} \\
\begin{tabular}{|r| *{13}{c} |c|}
	\hline
	\bm{$M$} & \multicolumn{13}{c}{\bm{$x$}} \vline & \bm{$L$} \\ 
	&	0	&	0.0067	&	0.0133	&	0.0200	&	0.0267	&	0.0333	&	0.0400	&	0.04667	&	0.0533	&	0.0600	&	0.0667	&	0.0733	&	0.0800	&	\\
	\hline
	20	&	384	&	384	&	384	&	384	&	384	&	384	&	400	&	400	&	400	&	400	&	400	&	400	&	400	&	5000000	\\
	40	&	224	&	208	&	208	&	208	&	208	&	208	&	208	&	208	&	208	&	208	&	208	&	208	&	208	&	5000000	\\
	60	&	208	&	208	&	208	&	208	&	208	&	208	&	208	&	208	&	208	&	208	&	208	&	208	&	208	&	5000000	\\
	80	&	208	&	208	&	208	&	208	&	208	&	208	&	208	&	208	&	208	&	208	&	208	&	208	&	208	&	5000000	\\
	100	&	208	&	208	&	208	&	208	&	208	&	208	&	208	&	208	&	208	&	208	&	208	&	208	&	208	&	5000000	\\
	120	&	208	&	208	&	208	&	208	&	208	&	208	&	208	&	208	&	208	&	208	&	208	&	208	&	208	&	5000000	\\
	140	&	176	&	240	&	240	&	256	&	272	&	272	&	272	&	256	&	240	&	272	&	224	&	256	&	192	&	5000000	\\
	160	&	208	&	208	&	208	&	208	&	208	&	208	&	208	&	208	&	208	&	208	&	208	&	208	&	208	&	5000000	\\
	180	&	256	&	288	&	288	&	288	&	272	&	272	&	272	&	272	&	272	&	272	&	288	&	288	&	288	&	5000000	\\
	200	&	416	&	416	&	400	&	416	&	400	&	368	&	384	&	400	&	400	&	384	&	400	&	368	&	416	&	5000000	\\
	\hline
\end{tabular} \\

\vspace{2ex} 
\pagebreak
\noindent
\bm{$p=0.25$} \\
\begin{tabular}{|r| *{13}{c} |c|}
	\hline
	\bm{$M$} & \multicolumn{13}{c}{\bm{$x$}} \vline & \bm{$L$} \\ 
	&	0	&	0.0067	&	0.0133	&	0.0200	&	0.0267	&	0.0333	&	0.0400	&	0.04667	&	0.0533	&	0.0600	&	0.0667	&	0.0733	&	0.0800	&	\\
	\hline
	20	&	421	&	421	&	421	&	421	&	421	&	416	&	416	&	416	&	416	&	416	&	416	&	416	&	416	&	5000000	\\
	40	&	390	&	390	&	390	&	395	&	390	&	390	&	395	&	400	&	395	&	390	&	400	&	400	&	395	&	5000000	\\
	60	&	400	&	400	&	400	&	400	&	400	&	400	&	400	&	400	&	400	&	400	&	400	&	400	&	400	&	5000000	\\
	80	&	354	&	356	&	510	&	360	&	360	&	208	&	360	&	360	&	356	&	356	&	358	&	356	&	358	&	5000000	\\
	100	&	624	&	624	&	624	&	624	&	624	&	624	&	624	&	624	&	624	&	624	&	624	&	624	&	624	&	5000000	\\
	120	&	384	&	384	&	400	&	368	&	384	&	384	&	368	&	384	&	384	&	368	&	368	&	368	&	384	&	5000000	\\
	140	&	342	&	346	&	470	&	346	&	344	&	208	&	352	&	336	&	352	&	346	&	348	&	332	&	346	&	5000000	\\
	160	&	366	&	368	&	754	&	378	&	372	&	1096	&	380	&	372	&	380	&	380	&	382	&	384	&	376	&	5000000	\\
	180	&	416	&	416	&	416	&	416	&	416	&	416	&	416	&	416	&	416	&	416	&	416	&	416	&	400	&	5000000	\\
	\hline
\end{tabular} \\

%\pagebreak
\noindent
\bm{$p=0.3$} \\
\begin{tabular}{|r| *{13}{c} |c|}
	\hline
	\bm{$M$} & \multicolumn{13}{c}{\bm{$x$}} \vline & \bm{$L$} \\ 
	&	0	&	0.0067	&	0.0133	&	0.0200	&	0.0267	&	0.0333	&	0.0400	&	0.04667	&	0.0533	&	0.0600	&	0.0667	&	0.0733	&	0.0800	&	\\
	\hline
	40	&	208	&	208	&	208	&	208	&	208	&	208	&	208	&	208	&	208	&	208	&	208	&	208	&	208	&	5000000	\\
	60	&	208	&	208	&	208	&	208	&	208	&	208	&	208	&	208	&	208	&	208	&	208	&	208	&	208	&	5000000	\\
	80	&	208	&	208	&	208	&	208	&	208	&	208	&	208	&	208	&	416	&	208	&	208	&	208	&	208	&	5000000	\\
	100	&	416	&	416	&	416	&	416	&	416	&	416	&	416	&	416	&	416	&	416	&	416	&	416	&	416	&	5000000	\\
	120	&	358	&	356	&	508	&	356	&	354	&	208	&	358	&	356	&	358	&	358	&	360	&	360	&	356	&	5000000	\\
	140	&	208	&	208	&	208	&	208	&	208	&	208	&	208	&	208	&	208	&	208	&	208	&	208	&	208	&	5000000	\\
	160	&	288	&	292	&	408	&	276	&	280	&	448	&	282	&	294	&	292	&	292	&	296	&	288	&	288	&	5000000	\\
	180	&	192	&	192	&	192	&	192	&	176	&	176	&	176	&	176	&	160	&	176	&	192	&	192	&	192	&	5000000	\\
	\hline
\end{tabular} \\

\vspace{2ex} 
%\pagebreak
\noindent
\bm{$p=1/3$} \\
\begin{tabular}{|r| *{13}{c} |c|}
	\hline
	\bm{$M$} & \multicolumn{13}{c}{\bm{$x$}} \vline & \bm{$L$} \\ 
	&	0	&	0.0067	&	0.0133	&	0.0200	&	0.0267	&	0.0333	&	0.0400	&	0.04667	&	0.0533	&	0.0600	&	0.0667	&	0.0733	&	0.0800	&	\\
	\hline
	20	&	624	&	624	&	624	&	624	&	624	&	624	&	624	&	624	&	624	&	624	&	624	&	624	&	624	&	5000000 \\
	40	&	624	&	624	&	624	&	624	&	624	&	624	&	624	&	624	&	624	&	624	&	624	&	624	&	624	&	5000000 \\
	60	&	632	&	632	&	632	&	632	&	632	&	632	&	632	&	632	&	632	&	632	&	632	&	632	&	632	&	5000000 \\
	80	&	624	&	625	&	625	&	630	&	624	&	624	&	625	&	630	&	640	&	630	&	640	&	655	&	624	&	5000000 \\
	100	&	624	&	624	&	624	&	624	&	624	&	640	&	624	&	624	&	624	&	624	&	624	&	624	&	824	&	5000000 \\
	120	&	624	&	624	&	624	&	624	&	624	&	624	&	624	&	624	&	624	&	624	&	624	&	624	&	624	&	5000000 \\
	140	&	624	&	624	&	624	&	624	&	624	&	624	&	624	&	624	&	624	&	624	&	624	&	624	&	624	&	5000000 \\
	160	&	624	&	624	&	624	&	624	&	624	&	624	&	624	&	624	&	624	&	624	&	624	&	624	&	624	&	5000000 \\
	180	&	624	&	624	&	640	&	624	&	624	&	624	&	640	&	640	&	624	&	624	&	624	&	624	&	624	&	5000000 \\
	200	&	624	&	624	&	624	&	624	&	624	&	624	&	624	&	624	&	624	&	624	&	624	&	624	&	624	&	5000000 \\
	\hline
\end{tabular} \\

\vspace{2ex}
\pagebreak
\noindent	
\bm{$p=0.35$} \\
\begin{tabular}{|r| *{13}{c} |c|}
	\hline
	\bm{$M$} & \multicolumn{13}{c}{\bm{$x$}} \vline & \bm{$L$} \\ 
	&	0	&	0.0067	&	0.0133	&	0.0200	&	0.0267	&	0.0333	&	0.0400	&	0.04667	&	0.0533	&	0.0600	&	0.0667	&	0.0733	&	0.0800	&	\\
	\hline
	20	&	208	&	208	&	192	&	208	&	192	&	208	&	208	&	192	&	208	&	208	&	192	&	208	&	208	&	5000000	\\
	40	&	208	&	208	&	208	&	208	&	208	&	208	&	208	&	208	&	208	&	208	&	208	&	208	&	208	&	5000000	\\
	60	&	208	&	208	&	208	&	208	&	208	&	208	&	208	&	208	&	208	&	192	&	192	&	208	&	192	&	5000000	\\
	80	&	208	&	192	&	208	&	208	&	208	&	208	&	192	&	208	&	208	&	192	&	208	&	208	&	208	&	5000000	\\
	100	&	192	&	192	&	176	&	192	&	176	&	192	&	144	&	176	&	192	&	208	&	192	&	160	&	192	&	5000000	\\
	120	&	208	&	208	&	192	&	208	&	208	&	208	&	208	&	208	&	208	&	192	&	208	&	208	&	208	&	5000000	\\
	140	&	208	&	208	&	208	&	208	&	192	&	208	&	192	&	208	&	208	&	208	&	208	&	208	&	208	&	5000000	\\
	160	&	208	&	176	&	192	&	160	&	192	&	192	&	192	&	176	&	176	&	176	&	176	&	160	&	208	&	5000000	\\
	180	&	400	&	416	&	416	&	416	&	416	&	384	&	416	&	416	&	416	&	416	&	416	&	416	&	416	&	5000000	\\
	200	&	208	&	208	&	208	&	208	&	208	&	208	&	208	&	208	&	208	&	208	&	208	&	208	&	208	&	5000000	\\
	\hline
\end{tabular} \\

\vspace{2ex} 
\noindent	
\bm{$p=0.36$} \\
\begin{tabular}{|r| *{13}{c} |c|}
	\hline
	\bm{$M$} & \multicolumn{13}{c}{\bm{$x$}} \vline & \bm{$L$} \\ 
	&	0	&	0.0067	&	0.0133	&	0.0200	&	0.0267	&	0.0333	&	0.0400	&	0.04667	&	0.0533	&	0.0600	&	0.0667	&	0.0733	&	0.0800	&	\\
	\hline
	20	&	192	&	176	&	192	&	144	&	160	&	192	&	192	&	192	&	192	&	208	&	208	&	192	&	160	&	5000000	\\
	40	&	208	&	208	&	208	&	208	&	208	&	208	&	208	&	208	&	208	&	208	&	208	&	208	&	208	&	5000000	\\
	60	&	208	&	208	&	208	&	208	&	208	&	208	&	208	&	192	&	208	&	208	&	208	&	208	&	208	&	5000000	\\
	80	&	208	&	208	&	208	&	208	&	208	&	208	&	208	&	208	&	208	&	192	&	208	&	208	&	208	&	5000000	\\
	120	&	208	&	208	&	208	&	208	&	208	&	208	&	208	&	208	&	208	&	208	&	208	&	208	&	208	&	5000000	\\
	160	&	208	&	208	&	208	&	208	&	208	&	208	&	208	&	208	&	208	&	208	&	208	&	208	&	208	&	5000000	\\
	200	&	208	&	192	&	192	&	208	&	208	&	208	&	208	&	208	&	192	&	192	&	208	&	208	&	192	&	5000000	\\
	\hline
\end{tabular} \\

\vspace{2ex}
%\pagebreak
\noindent	
\bm{$p=0.4$} \\
\begin{tabular}{|r| *{13}{c} |c|}
	\hline
	\bm{$M$} & \multicolumn{13}{c}{\bm{$x$}} \vline & \bm{$L$} \\ 
	&	0	&	0.0067	&	0.0133	&	0.0200	&	0.0267	&	0.0333	&	0.0400	&	0.04667	&	0.0533	&	0.0600	&	0.0667	&	0.0733	&	0.0800	&	\\
	\hline
	20	&	416	&	416	&	416	&	416	&	416	&	416	&	416	&	416	&	416	&	416	&	416	&	416	&	416	&	5000000 \\
	40	&	416	&	416	&	416	&	416	&	416	&	416	&	416	&	416	&	416	&	416	&	416	&	416	&	416	&	5000000 \\
	60	&	416	&	416	&	416	&	416	&	416	&	416	&	416	&	416	&	416	&	416	&	416	&	416	&	416	&	5000000 \\
	80	&	208	&	208	&	208	&	208	&	208	&	208	&	208	&	208	&	208	&	208	&	208	&	208	&	208	&	5000000 \\
	100	&	208	&	208	&	208	&	208	&	208	&	208	&	208	&	208	&	208	&	208	&	208	&	208	&	208	&	5000000 \\
	120	&	208	&	208	&	208	&	208	&	208	&	208	&	208	&	208	&	208	&	208	&	208	&	208	&	208	&	5000000 \\
	140	&	208	&	208	&	208	&	208	&	208	&	208	&	208	&	208	&	208	&	208	&	208	&	208	&	208	&	5000000 \\
	160	&	208	&	208	&	208	&	208	&	208	&	208	&	208	&	208	&	208	&	208	&	208	&	208	&	208	&	5000000 \\
	180	&	208	&	208	&	208	&	192	&	208	&	192	&	208	&	192	&	208	&	208	&	208	&	208	&	208	&	5000000 \\
	\hline
\end{tabular} \\

%\end{widetext}
\twocolumngrid

%\bibliography{RN2-bibliography}
%\bibliographystyle{apsrev4-1}

\end{document}